\documentclass[fleqn,usenatbib]{mnras}

\usepackage{newtxtext,newtxmath}
\usepackage[T1]{fontenc}

\DeclareRobustCommand{\VAN}[3]{#2}
\let\VANthebibliography\thebibliography
\def\thebibliography{\DeclareRobustCommand{\VAN}[3]{##3}\VANthebibliography}

\usepackage{graphicx}	% Including figure files
\usepackage{amsmath}	% Advanced maths commands

\defcitealias{2006AJ....131..571H}{H06}
\defcitealias{2011MNRAS.417...93R}{RH11}
\defcitealias{2017AJ....153...10M}{M17}
\defcitealias{2019MNRAS.482..715L}{L19}
\defcitealias{2021MNRAS.501.5149Q}{Q21}
\defcitealias{2021MNRAS.502.1753T}{T21}
\defcitealias{2021A&A...649A...6G}{G21}

\title[Exploring faint WDs and LF with HSC]{Exploring faint white dwarfs and the luminosity
function with Subaru HSC and SDSS in Stripe 82}

\author[Tian Qiu et al.]{
Tian~Qiu$^{1,2,3}$\thanks{E-mail: tian.qiu@ipmu.jp},
Masahiro~Takada$^{1,3}$,
Naoki~Yasuda$^{1}$,
Akira~Tokiwa$^{1,2,3}$,
Kazumi~Kashiyama$^{4,1}$,
\and
Yoshihisa~Suzuki$^{4}$,
Kenta~Hotokezaka$^{5,1}$
\\
$^{1}$ Kavli Institute for the Physics and Mathematics of the Universe (WPI), The University of Tokyo Institutes for Advanced Study (UTIAS), 
\\The University of Tokyo, 5-1-5 Kashiwanoha, Kashiwa-shi, Chiba, 277-8583, Japan
\\
$^{2}$ Department of Physics, The University of Tokyo, 7-3-1 Hongo, Bunkyo-ku, Tokyo 113-0033 Japan\\
$^{3}$ Center for Data-Driven Discovery (CD3), Kavli IPMU (WPI), UTIAS, The University of Tokyo, Kashiwa-shi, Chiba 277-8583, Japan\\
$^{4}$ Astronomical Institute, Graduate School of Science, Tohoku University, Aoba, Sendai 980-8578, Japan\\
$^{5}$  Research Center for the Early Universe, Graduate School of Science, University of Tokyo, Bunkyo-ku, Tokyo 113-0033, Japan
}

\date{Accepted 2024 October 10. Received 2024 October 10; in original form 2024 January 31}

\pubyear{2024}

\begin{document}
\label{firstpage}
\pagerange{\pageref{firstpage}--\pageref{lastpage}}
\maketitle

\begin{abstract}
We present 5,080 white dwarf (WD) candidates selected from stars matching between the multi-band imaging datasets of the Subaru Hyper Suprime-Cam (HSC) Survey and the Sloan Digital Sky Survey (SDSS) in the Stripe82 region covering about 165~deg$^2$. 
We select WD candidates from the ``reduced proper motion'' diagram by combining the apparent magnitude in the range $i=19$ -- 24 and the proper motion measured from the datasets among a baseline of $\sim$ 14~years.
We refine the WD candidates by fitting blackbody and template WD atmosphere models to HSC photometries for each candidate,
enabling the estimation of distance and tangential velocity ($v_{\rm t}$).
The deep HSC data allow us to identify low-temperature ($<4000$~K) and faint WD candidates down to absolute magnitude, $M_{\rm bol}\simeq 17$.
We evaluate the selection function of our WD candidates using a mock catalogue of spatial and kinematic distributions of WDs in the (thin and thick) disc and halo regions based on a Galactic model. 
We construct samples of disc and halo WD candidates by selecting WDs with tangential velocity, $40<v_{\rm t}/[{\rm km}~{\rm s}^{-1}]<80$ and $200<v_{\rm t}/[{\rm km}~{\rm s}^{-1}]<500$, respectively.
The total number densities of the disc and halo WDs
are $(9.33 \pm 0.89) \times 10^{-3}$~pc$^{-3}$ and $(6.34 \pm 2.90) \times 10^{-4}$~pc$^{-3}$.
Our luminosity functions (LF) extend down to fainter absolute magnitudes compared with previous work.
The faint WDs could represent the oldest generation of building blocks 
over the past $\sim$10 billion years of the assembly history of our Milky Way.
\end{abstract}


%%%%%%%%%%%%%%%%%%%%%%%%%%%%%%%%%%%%%%%%%%%%%%%%%%

%%%%%%%%%%%%%%%%% BODY OF PAPER %%%%%%%%%%%%%%%%%%

\section{Introduction}
\label{sec:introduction}

White dwarfs (WDs) are the remnants of low- to intermediate-mass (lighter than 7-11 M$_\odot$, depending on their metallicity) stars that have exhausted their nuclear fuel \citep{1999ApJ...515..381R,2008MNRAS.387.1693C,2013ApJ...765L..43I,2015ApJ...810...34W,2018MNRAS.480.1547L}.
As these objects cool and fade over billions of years, they traverse a well-defined path in the colour-magnitude diagram, known as the WD cooling sequence \citep{2010ApJ...717..183R}. 
The distribution of WDs, as quantified by the luminosity function, encodes valuable information to study our Milky Way Galaxy.

There are many previous works on statistics of WDs. For instance, \citet{2016NewAR..72....1G} showed the white dwarf luminosity function (WDLF) is a fundamental tool for studying the properties of the WD population. 
The position and shape of the peak and turnover in the luminosity function at the faint end provide valuable information about how the oldest population has formed in the assembly history of Milky Way \citep{1979ApJ...233..226L, 1987ApJ...315L..77W, 2013MNRAS.434.1549R}.

The challenge of probing the turnover of the luminosity function lies in the presence of faint WDs that exhibit colours similar to subdwarfs. 
This made it difficult to distinguish them from the subdwarf population, leading to uncertainties in understanding the turnover. 
The breakthrough came with the work of \cite{1988ApJ...332..891L}, who successfully resolved the peak and identified WD samples fainter than the turnover. 
Subsequent studies have made significant contributions to the understanding of the WDLF.
Earlier investigations, such as the studies by \cite{1992MNRAS.255..521E}, \cite{1996Natur.382..692O}, \cite{2013ApJ...765L..43I}, and \cite{1999MNRAS.306..736K}, focused on smaller WD samples with proper motion measurements, providing valuable insights into the WDLF and its relation to the Galactic disc. 
\citet{2005ApJS..156...47L} made a comprehensive work about the formation rate, mass and luminosity functions of WDs with hydrogen-dominated atmospheres (DA WDs).
The work by \citet[hereafter \citetalias{2006AJ....131..571H}]{2006AJ....131..571H} marked a milestone by increasing the sample size through the use of data from the SDSS \citep{1998AJ....116.3040G,2006AJ....131.2332G}.
They employed the reduced proper motion (RPM) selection method, combining SDSS and USNO-B \citep{2003AJ....125..984M} astrometry, to obtain a sample of 6,000 WDs. 
The following works based on SDSS spectra by \citet{2008AJ....135....1D} and \citet{2009A&A...508..339K} have also derived the luminosity functions for the bright segments.
Building upon this, the work by \citet[hereafter \citetalias{2011MNRAS.417...93R}]{2011MNRAS.417...93R} utilised a similar RPM approach to generate a sample of over 10,000 WDs from the SuperCOSMOS sky survey \citep{2001MNRAS.326.1279H,2001MNRAS.326.1295H,2001MNRAS.326.1315H}. 
Both of these studies extended their investigations beyond the Galactic disc to include the stellar halo.
\citet{2015MNRAS.450..743R} concentrated their research on DA WDs and compared observed distributions with simulations using data from the spectroscopic survey conducted by the Large Sky Area Multi-Object Fiber Spectroscopy Telescope \citep[LAMOST, ][]{2012RAA....12.1197C}, LAMOST Spectroscopic Survey of the Galactic Anti-center \citep[LSS-GAC, ][]{2015MNRAS.448..855Y}, while they also investigated the mass function of hydrogen-rich WDs from SDSS in \citet{2015MNRAS.452.1637R}.
The work by \citet[hereafter \citetalias{2017AJ....153...10M}]{2017AJ....153...10M} delved deeper into the WD analysis in both the disc and halo populations. 
Further advancements were made by \citet[hereafter \citetalias{2019MNRAS.482..715L}]{2019MNRAS.482..715L}, including a larger number of faint WDs and employed more sophisticated methods, utilising data from the Panoramic Survey Telescope \& Rapid Response System \citep[Pan-STARRS, ][]{2010SPIE.7733E..0EK,2016arXiv161205560C}, to recover the luminosity function with improved accuracy and precision.
The works by \citet[hereafter \citetalias{2021MNRAS.502.1753T}]{2021MNRAS.502.1753T} and \citet[hereafter \citetalias{2021A&A...649A...6G}]{2021A&A...649A...6G}, based on {\it Gaia} Data Release 2 \citep[DR2, ][]{2018A&A...616A...1G} and Early Data Release 3 \citep[EDR3, ][]{2021A&A...649A...1G} respectively, have constructed the luminosity function using a complete sample within 100~pc, 
which have yielded an unprecedented well-characterised catalogue in the solar neighbourhood, offering a comprehensive insight into nearby WDs.
These efforts have contributed significantly to our understanding of the WDLF and the properties of WDs in different Galactic components.

Constructing the WDLF is not a straightforward task due to various observational biases and limitations. 
For instance, WDs are faint objects, making them difficult to detect at large distances. 
Furthermore, the detection efficiency can vary with colour, as WDs become redder and fainter as they cool. 
Therefore, careful consideration must be given to these factors when deriving the WDLF from observational data.
There are numerous statistical methods available for estimating the luminosity function.
Among these, the maximum volume density estimator \citep{1968ApJ...151..393S} is widely used in many studies of the WD luminosity function such as \citetalias{2006AJ....131..571H}, \citetalias{2011MNRAS.417...93R}, \citet{2015MNRAS.450.4098L}, \citetalias{2017AJ....153...10M}, \citetalias{2019MNRAS.482..715L} and \citetalias{2021A&A...649A...6G}, attesting to its effectiveness and robustness.
Nevertheless, those prior studies have been limited by their shallow photometries, typically featuring a faint magnitude limit of approximately 20 mags. 
This limitation restricts our understanding of faint white dwarfs.
Hence the purpose of this paper is to construct a catalogue of WDs down to fainter magnitudes by taking full advantage of a significantly deeper survey, the Hyper Suprime-Cam Subaru Strategic Program \citep[HSC-SSP, ][]{2018PASJ...70S...4A}. 
To do this we combine HSC data with
the Stripe~82 data of the SDSS,
which is the deepest among the SDSS data sets, 
to obtain the proper motion measurements of stars down to a magnitude of 24.
With precise astrometry and long-time baseline, we are able to further expand our capacity to trace those faint white dwarfs.
We also perform a fitting of the blackbody or WD atmosphere model to the high-quality HSC photometries for each WD candidate to refine the WD candidates and estimate the photometric distances. We will use this information to estimate the WDLF.

This paper is structured as follows. 
In Section \ref{sec:data}, we introduce the properties of Subaru HSC and SDSS Stripe~82 data and other datasets in this work.
The corresponding algorithm to process our catalogue is presented in Section~\ref{sec:Measurement}.
The selection criteria for the WD candidates are described in Section \ref{sec:selection}.
Section~\ref{sec:method} describes the methods to derive our luminosity function.
Section~\ref{sec:WDLF} presents our results of WDLFs and those interesting candidates of extremely faint WDs.
Finally, we summarise in Section~\ref{sec:conclusion}.

%%%%%%%%%%%%%%%%%%%%%%%%%%%%%%%%%%%%%%%%%%%%%%%%%%
\section{Data}
\label{sec:data}

\subsection{Subaru HSC-SSP}
\label{ssec:HSC}
The Hyper Suprime-Cam is a wide-field imaging camera on the prime focus of the 8.2-m Subaru Telescope that boasts an impressive field of view of 1.77 deg$^2$, making it an ideal instrument for large-scale sky surveys \citep{2018PASJ...70S...1M,2018PASJ...70S...2K, 2018PASJ...70S...4A, 
2018PASJ...70...66K,2018PASJ...70S...3F}.
The HSC-SSP has carried out a 330-night multi-band imaging survey of the high-latitude sky since 2014 \citep{2018PASJ...70S...8A}, providing a wealth of high-quality data.
The HSC data used in this work are taken from the Wide layer, which covers over $\sim$1100~deg$^2$ in five broad bands ($grizy$), with a 5$\sigma$ point-source depth of ($g$, $r$, $i$, $z$, $y$) $\simeq$ (26.5, 26.1, 26.2, 25.1, 24.4). 
The total exposure times range from 10~min for $g$, $r$-bands to 20~min for $i$, $z$, $y$-bands.
We use the dataset from the internal data release labelled as `S21A', which is based on the data observed from March 2014 to January 2021.
The data was reduced using
the same pipeline as that in Data Release~3 \citep[DR3, ][]{2022PASJ...74..247A}.

As mentioned in \cite{2022PASJ...74..247A}, $i$-band images are preferentially taken when the seeing is good.
The $i$-band images have a median point spread function (PSF) full width at half maximum of $0.6\arcsec$. 
The superb image quality of the HSC $i$-band images has significant benefits for our study. 
It allows for a robust selection of stars and minimises contamination from galaxies in the star catalogue. 
This is crucial for accurately identifying and characterising WD candidates in our analysis. 
To ensure the accuracy and reliability of the data reduction process, the software applied in the pipeline is described in \cite{2018PASJ...70S...5B} and \cite{2018PASJ...70S...3F}. 
These papers provide detailed information about the data reduction procedures.
The photometric and astrometric calibration of the HSC data was performed using the publicly available Pan-STARRS catalogue \citep[PS, ][]{2012ApJ...756..158S,2012ApJ...750...99T,2013ApJS..205...20M,2016arXiv161205560C,2020ApJS..251....6M}. 

In this work, our analysis is based on the primary photometric sources with $i_{\rm HSC}<24.5$.
The master HSC catalogue used in our study contains approximately 13.4~million objects, including 1.45~million stars.
To ensure the completeness of our analysis, we prefer an overqualified sample to include more potential objects.
We will describe the selection of our WD candidates in the following.

\subsection{SDSS Stripe82}
\label{ssec:S82}
The SDSS utilised a dedicated wide-field 2.5m telescope, specifically designed to image the sky \citep{2000AJ....120.1579Y,2006AJ....131.2332G}. 
The survey was conducted in drift-scanning mode, capturing images of the sky across five optical filters, namely $ugriz$ \citep{1996AJ....111.1748F,1998AJ....116.3040G}.

In this work, we are using a special stripe region of  
SDSS that
imaged a 275 deg$^2$ region multiple times around the Celestial Equatorial region, known as Stripe82 in Data Release~7 (DR7, hereafter S82 referring to the SDSS dataset). 
The coadd images in $ugriz$ are approximately 1.5~magnitudes deeper than those in the SDSS single-pass data \citep{2014ApJ...794..120A,2014ApJS..213...12J}.
This region is $2.5^{\circ}$ wide stripe along the Celestial Equator in the South Galactic Cap, more precisely the contiguous region in the ranges of $-50^{\circ}\le {\rm R.A.}\le 60^{\circ}$ and $-1.25^{\circ}\le {\rm Dec.}\le 1.25^{\circ}$. 
The publicly-available data and catalogues of S82 are constructed from the co-adds of approximately 20 multiple images, taken from September 1998 to November 2005. 
The median seeing is 1.1$\arcsec$ and the 50\% completeness limits are $r=23.5$ for galaxies and $r=24.3$ for stars, respectively.

There are approximately 165~deg$^2$ of the S82 footprint overlapping with the HSC S21A data,
which provides about 8 million objects with $r<24$ in total. 
For the positions of stars/galaxies, we use their centroid positions in the HSC $i$-band as our fiducial choice.
We will also use the centroid positions of the other filters to test the possible effects of systematic errors on our results.
However, we will not adopt any SDSS measurements, including the photometries and the positions, to our analysis after matching.

\subsection{Other datasets}
\label{ssec:otherdata}
We also use {\it Gaia}\footnote{\url{https://www.cosmos.esa.int/web/gaia/data-release-3}} DR3 \citep{2023A&A...674A...1G} stars and 
SDSS DR16 quasars\footnote{\url{https://www.sdss4.org/dr17/algorithms/qso_catalog/}}
\citep{2020ApJS..250....8L} for calibration and validation.
In this work, there are around 0.2 million stars in {\it Gaia} matched to our star catalogue to do astrometric correction caused by the differential chromatic refraction (DCR), 
which is an optical effect that causes the light of different colours (or wavelengths) to be refracted by different amounts as it passes through the atmosphere \citep{1982PASP...94..715F}.
The quasars are used to estimate statistical errors of the proper motion measurements.
In addition, we use the SDSS quasars to validate our correction method of the DCR effect
\citep[see Appendix~D in][hereafter \citetalias{2021MNRAS.501.5149Q}, for details]{2021MNRAS.501.5149Q}.

%%%%%%%%%%%%%%%%%%%%%%%%%%%%%%%%%%%%%%%%%%%%%%%%%%
\section{Measurements}
\label{sec:Measurement}
We follow the method developed in \citetalias{2021MNRAS.501.5149Q} to do both photometric and astrometric calibration, to perform the proper motion measurements. 
In this paper, we use the improved matching algorithm to include larger proper-motion objects and minimise the contamination.
We will describe the details of these methods in the following sections.

\subsection{Photometric calibration}
\label{ssec:PhotoCali}
We use the photometries and colours in the HSC catalogue for each object throughout this paper unless otherwise stated.
However, we must still adopt the SDSS photometry in our matching algorithm to screen out the mismatched pairs. 
The HSC filter system (transmission curve, wavelength coverage, etc.) is quite similar but not exactly the same as the SDSS system.
Therefore, we need to infer the SDSS-like photometry of HSC from the current colour-magnitude relationship. 

\begin{figure}
\begin{center}
 \includegraphics[width=\columnwidth]{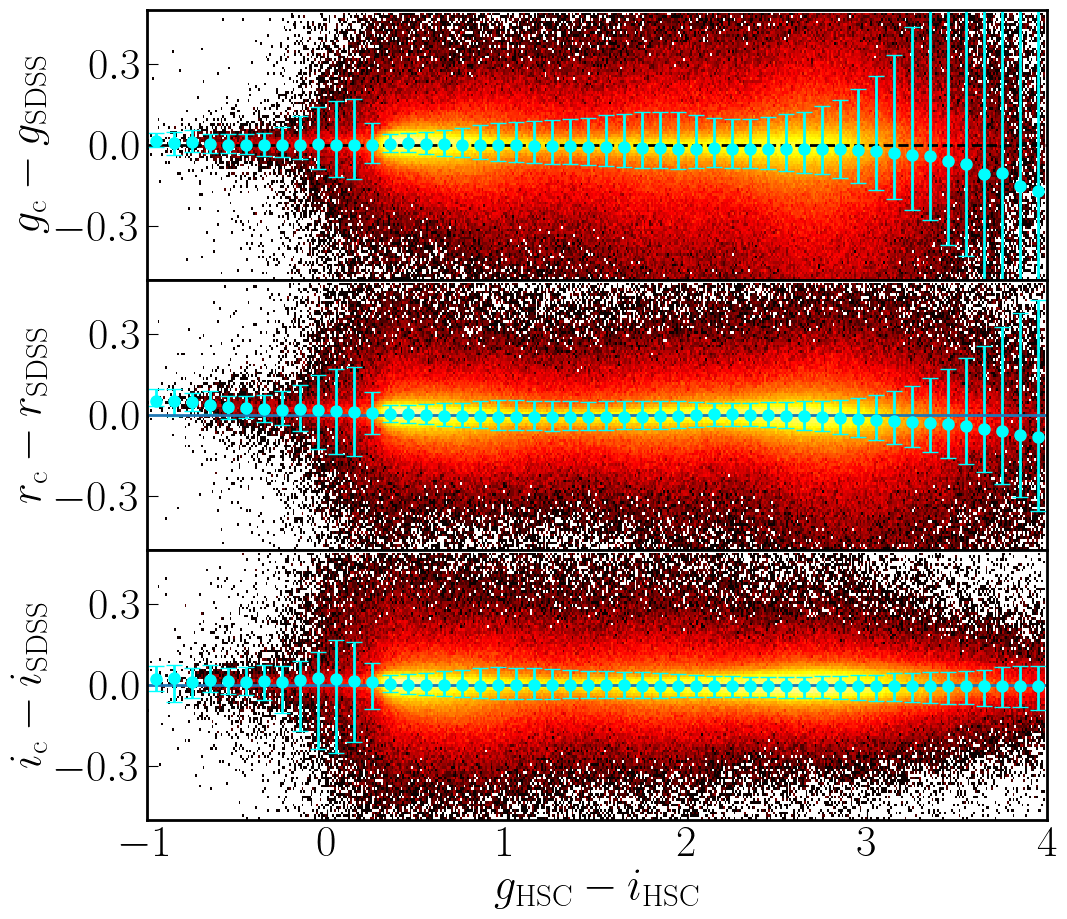}
 \caption{The magnitude difference of $g,r,i$-bands between HSC after photometric calibration ($x_\mathrm{c}$) and SDSS ($x_\mathrm{SDSS}$) against the colour $g-i$ in the HSC filter system. 
 The cyan points with errorbars show the median with 1$\sigma$ error in each 0.1~mag colour bin.
 }
\label{fig:photometriccali}
\end{center}
\end{figure}

Following the work in \citetalias{2021MNRAS.501.5149Q}, we transfer the HSC PSF magnitude into SDSS-like photometry by employing the second-order polynomial models fitted with the magnitude difference between HSC and SDSS ($g-g_\mathrm{SDSS}$, $r-r_\mathrm{SDSS}$, $i-i_\mathrm{SDSS}$) against the HSC colour ($g-i$, $g-i$, $r-i$) for main-sequence star objects, as shown in Figure \ref{fig:photometriccali}.
The median magnitude difference in each colour bin is almost zero for $g-i$ between -0.5 and 3.
The 1$\sigma$ error is less than 0.1~mag for $r,i$-bands in the dense region, while the error for $g$-band is around 0.1~mag.
Even at the extreme blue and red ends of the colour, the $i$-band photometry exhibits excellent consistency, while the $g$ and $r$-bands also display acceptable agreement within the majority of the colour range.
The deviation at the red end can be attributed to the limited number of samples in this region. 
The deviation is not of significant concern for our study, as our primary interest lies in the bluer parts where WDs are predominantly located. 
Even the faintest WDs are expected only to reach a $g-i$ colour index of approximately 2.
We apply this transformation during the matching algorithm process to filter out mismatched pairs, as detailed in Section~\ref{ssec:MatchAlgo}, and during the completeness analysis in Section~\ref{ssec:completeness}. 
 
\subsubsection{Reddening correction}
\label{sssec:red}
In addition to the transformation between the HSC and SDSS filter systems, interstellar reddening affects the photometry as well. 
The observed colour of a star is a convolution of its intrinsic colour and the reddening effect imposed by the interstellar medium.

We adopt a two-step correction method.
Firstly, we correct for the interstellar extinction using the reddening maps of \cite{2015ApJ...810...25G}\footnote{\url{http://argonaut.skymaps.info}}.
It provides a reddening map as a function of the R.A. and Dec. positions, similarly to that in \citet[SFD]{1998ApJ...500..525S}.
For the S82 region, the dust extinction does not significantly vary with directions;
the reddening $E(B-V)$ is typically around 0.02~mag.
Hence we ignore the R.A. and Dec. dependence of the reddening for simplicity. 
Note that this reddening amount compared to the SFD, 
the ratio $R_{a-b}=E(a-b)/E(B-V)_{\mathrm{SFD}}$ is around 0.98 for most of the $g-r$ colour range. 
For the transformation from reddening to extinction for each band, we adopt the value from Table~6 of \cite{2011ApJ...737..103S}.
Since the HSC filter system is close to the PS filter system \citep{2010SPIE.7733E..0EK}, we use the column of $R_v=3.1$ of PS $grizy$ in this work, as given by Table~\ref{tab:red}.
Note that the relative extinctions between the different filters are based on the reddening law by \cite{1999PASP..111...63F}.

\begin{table}
 \begin{center}
 \begin{tabular}{lccccc}
  \hline
  \hline 
 Passband ($x$) & $g$ & $r$ & $i$ & $z$ & $y$\\[2pt]
 \hline
 $A_x/E(B-V)_{\mathrm{SFD}}$ & 3.172 & 2.271 & 1.682 &  1.322 & 1.087\\[2pt]
  \hline
 \end{tabular}
 \end{center}
 \caption{The coefficients for extinction in each passband $x$ from reddening, $A_x/E(B-V)_{\mathrm{SFD}}$.
 }
 \label{tab:red}
\end{table}
Using the above extinction model, we corrected the photometries in HSC $grizy$ bands for all objects used in this paper.

Based on the photometric distance we derived in Section~\ref{ssec:VmaxDist}, we refined the reddening estimates using the 3D extinction map from {\it Stilism}\footnote{\url{https://stilism.obspm.fr/}} \citep{2014A&A...561A..91L,2017A&A...606A..65C,2018A&A...616A.132L}. 
This adjustment allows for a more precise determination of extinction.

\subsection{Matching algorithm}
\label{ssec:MatchAlgo}
To measure the proper motion of each star object, we first need to perform a matching between objects in the HSC and 
S82 catalogues and then measure the proper motion from the angular offsets between the centroid positions over the time baseline (interval) in which the two datasets were taken (see \citetalias{2021MNRAS.501.5149Q}). 
In this paper, we further optimise the matching algorithm to include larger proper motion objects, minimising the mismatched pairs simultaneously.
The master HSC catalogue contains 1.4~million stars in the S82 region covering about 165~deg$^2$, while the SDSS S82 catalogue contains about 8 million objects (including stars and galaxies).

For faint objects in the S82 catalogue, the SDSS photometry is relatively unreliable, and the star/galaxy contamination can reach as high as 50\% at the faint end.
Given the fact that the HSC classification is secure, 
we match the HSC stars to all objects in S82 to avoid missing any potential pairs. 
Our searching radius is adjusted to within $2^{\prime\prime}$ in radius from the centre of each HSC star
to include high proper motion objects while avoiding too many contaminations. 
Although we could use a larger separate radius to find 
additional pairs, 
in this case, we would have too many multiple-matched objects and contaminated pairs for HSC stars.

\begin{figure}
\begin{center}
 \includegraphics[width=\columnwidth]{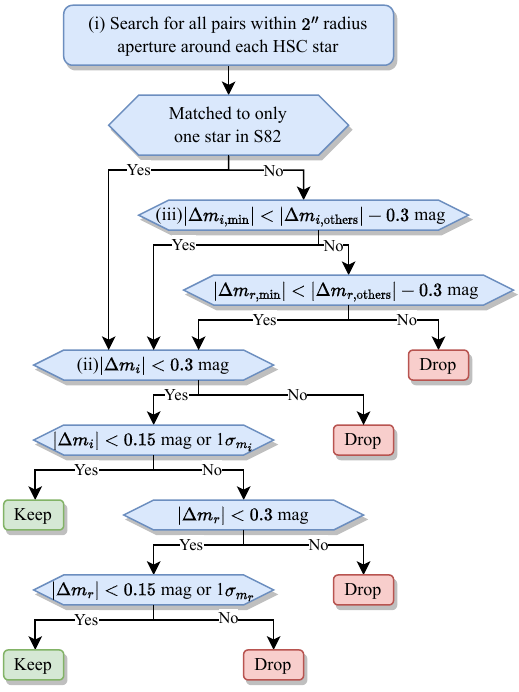}
 \caption{Flowchart of our matching algorithm. 
 In contrast to the selection of single-matched objects, multiple-matched candidates require additional verification steps.
 This requires that the magnitude difference of the chosen pair is at least 0.3~mag smaller than that of other pairs.
 }
\label{fig:flowchart}
\end{center}
\end{figure}

In addition to the search region for each star, 
we also apply matching of the magnitude for all the potential pairs to screen out the contaminations. 
Figure~\ref{fig:flowchart} shows a schematic flowchart of our matching algorithm:
\begin{enumerate}
    \item Search for all S82 objects to within $2^{\prime\prime}$ from the centre of each HSC star.
    If two or more objects in the S82 catalogue are matched to one HSC star, 
    we keep them as a potential pair and assign a label to indicate they are in the same group.
    \item We keep the single-matched pairs whose $i$-band magnitude offsets between HSC and S82 are less than 0.15~mag or 1$\sigma$ photometric error (up to 0.3~mag). 
    If the $i$-band offset is larger than 0.15~mag and 1$\sigma$ photometries but less than 0.3~mag, we check the
    $r$-band photometries. If the $r$-band magnitude difference is smaller than 0.15~mag or $1\sigma$ error, 
    we keep the pair as the matched pair. If the candidate does not pass these conditions, we discard it.
	\item[(iii-a)] For the multiple candidate pairs, we first keep a probable candidate that has the {\it smallest} $i$-band magnitude difference by at least $0.3$~mag  compared to other candidate pairs. 
	If any candidate pair does not meet this $i$-band condition, we check the $r$-band photometries and keep the pair, as the probable candidate, if the candidate has the smallest magnitude difference as the condition for the $i$-band difference. 
    If any candidate pair does not pass the $i$- or $r$-band magnitude difference conditions, we discard the candidate(s).
	\item[(iii-b)] For the candidate that passed the condition (iii-a), 
		we keep the pair if it has a $i$-band magnitude difference smaller than 0.15~mag or $1\sigma$.
		If this fails, we check the $r$-band magnitude difference and keep the pair if it passes the same condition as in the $i$-band. If the candidate pair does not pass the $i$- or $r$-band condition, we discard the candidate(s).
\end{enumerate}
Note that, if multiple HSC stars have the same S82 star(s) within their $2^{\prime\prime}$ search radius, 
we treat those as the same group of matched candidates. 
In addition, if one S82 star has multiple HSC candidates, we check the above conditions starting from the S82 star instead of an HSC star.
These sanity checks are mostly for the multiple candidate pairs (such groups).
After the above matching selection, we select 37.0\% objects from those multiple-matched pair candidates, 
while there are over 93.5\% single-matched pairs left.

\subsection{Proper motion calibration}
\label{ssec:PM}
We use the method in \citetalias{2021MNRAS.501.5149Q} to measure proper motions for the matched stars from the angular offsets over the time baseline.
Compared to \citetalias{2021MNRAS.501.5149Q}, we increased the matching radius to $2^{\prime\prime}$ from $1^{\prime\prime}$ and therefore can include stars that have larger proper motions.
Although we use matched galaxies in the HSC and S82 catalogues to define the reference frame for our proper motion measurements, 
residual systematic errors in the measurements do not relatively affect our main results from large proper-motion stars.

Besides, the {\it Gaia} stars are used for the correction of DCR as we mentioned in Section~\ref{ssec:otherdata}.
It can cause the apparent position offset of an astronomical object to shift depending on its colour and its altitude above the horizon.
Since the WDs cover a wider range of colours, we vary the colour $g-i$ from $-1$ to 4 to include bluer objects in addition to the correction.

It should be noted that we have not accounted for the subtraction of solar rotation motion in the analysis,
because we focus on the motion of WDs relative to our local rest frame, rather than the Galactic centre,
to identify WD candidates in the reduced proper motion diagram (see in Section~\ref{ssec:RPM}).

\subsubsection{Comparison with {\it Gaia}}
\label{ssec:comparisonwithgaia}
\begin{figure}
\begin{center}
 \includegraphics[width=\columnwidth]{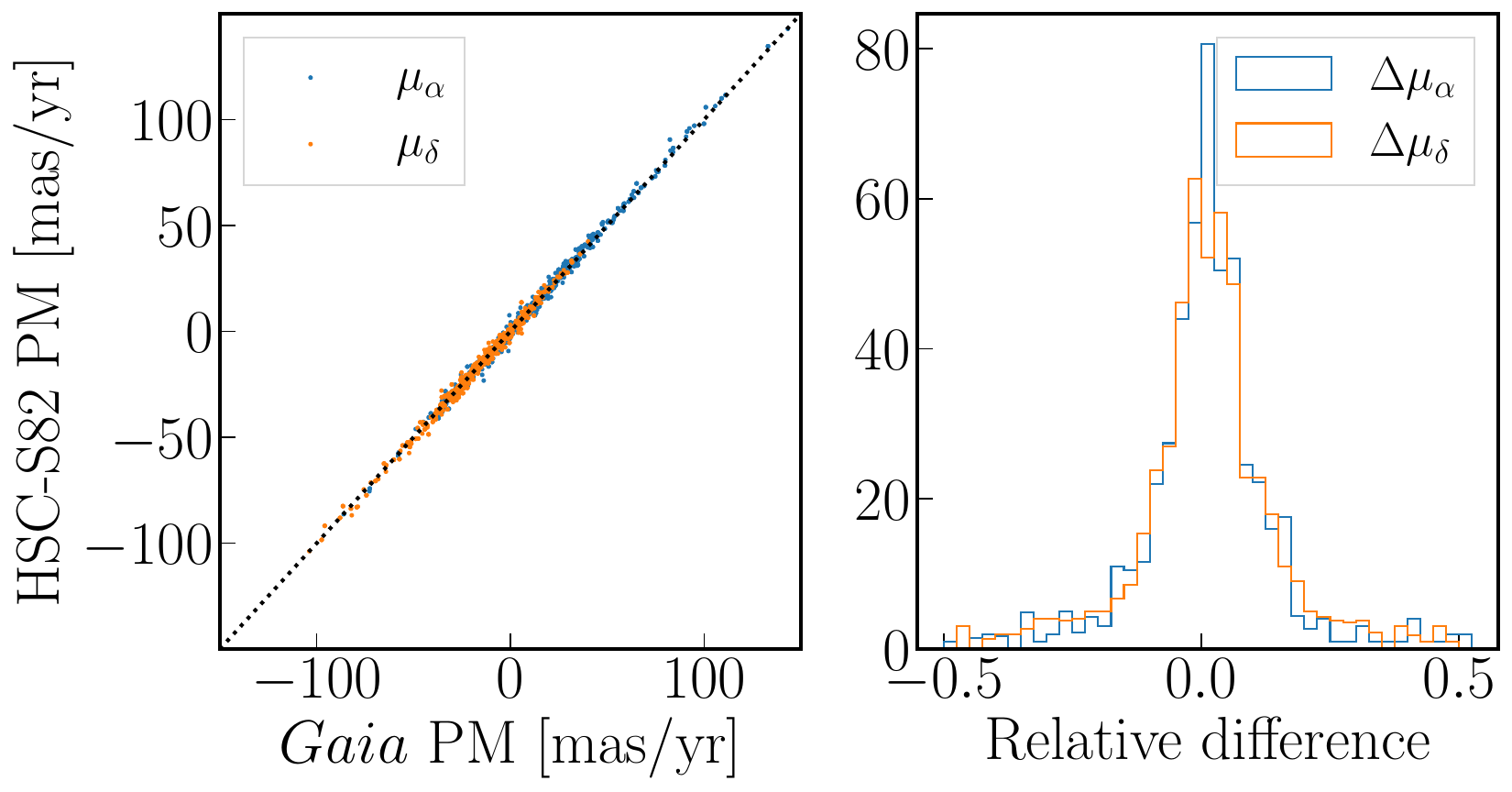}

 \caption{Comparison of proper motion measurements between the HSC-S82 catalogue and {\it Gaia}. 
 In the left panel, we display the proper motion measurements of matched objects from both HSC-S82 and {\it Gaia}. 
 The right panel presents a histogram of the relative differences in proper motion ($\Delta \mu= (\mu_{\mathrm{HSC-S82}}-\mu_{\it Gaia})/\mu_{\it Gaia}$). 
 The results in both the R.A. direction ($\mu_\alpha$) and the Dec. direction ($\mu_\delta$) show good consistency.
 }
\label{fig:Gaia_pm}
\end{center}
\end{figure}

To assess the accuracy of the proper motion measurements in our HSC-S82 catalogue, we compared them with those provided by the {\it Gaia}. 
We identified a total of 546 objects that are common to both datasets. 
As Figure~\ref{fig:Gaia_pm} shows, the results demonstrate good consistency in both R.A. and Dec. direction, confirming the reliability of the proper motion measurements in the HSC-S82 catalogue.
A more detailed comparison of the proper motion measurements between HSC-S82 and {\it Gaia} can be found in the appendix of \citetalias{2021MNRAS.501.5149Q}.

\subsection{Completeness}
\label{ssec:completeness}
The accuracy and reliability of the WD luminosity function are intrinsically linked to the completeness of the WD sample. 
An incomplete sample can result in a skewed or biased luminosity function, leading to inaccurate conclusions about the underlying WD population.
In this work, we identify our WD candidates using the reduced proper motion method, as described in Section~\ref{ssec:RPM}, which can introduce more difficulties and complexities in treating the completeness and contaminations.
For instance, if our matching algorithm fails to identify high-proper-motion WDs, the luminosity function could underestimate the density of intrinsically faint WDs. 
These stars are expected to be closer to us, and their omission could skew our understanding of the WD population.

Therefore, ensuring the completeness of the WD sample is a critical step in generating a robust WD luminosity function. 
The incompleteness primarily arises from two factors.
The first is the proper motion limits, which are dictated by the position matching and time baseline. 
During the matching process, high-proper-motion objects may be missed due to the limitation of the searching radius. 
To address this issue, we explore the upper limit of the proper motion for a complete sample, and this is discussed in detail in Section~\ref{sec:selection}.

\begin{figure}
\begin{center}
 \includegraphics[width=0.9\columnwidth]{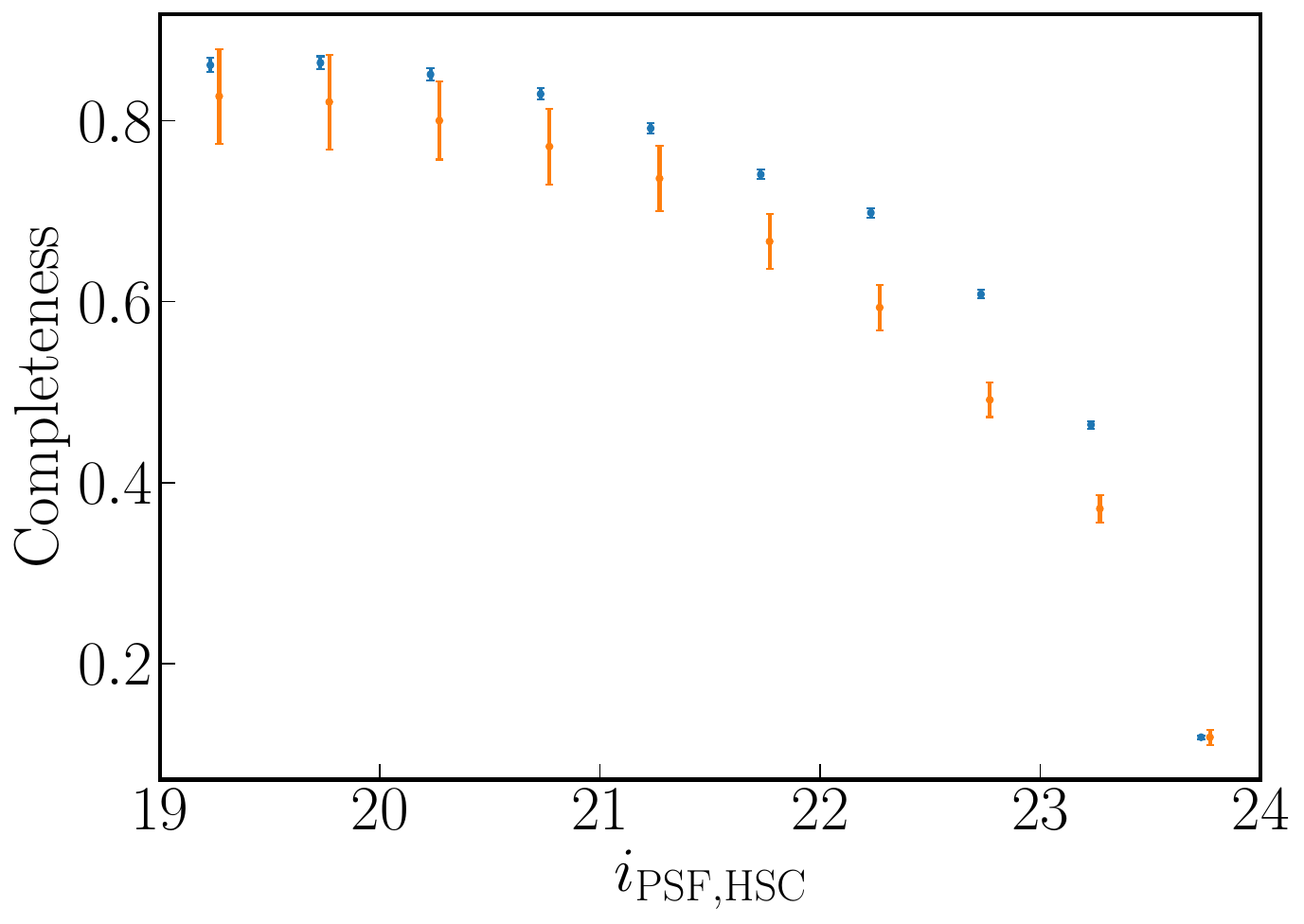}
 \caption{The completeness of the matched stars compared to the HSC identified stars against the $i$-band PSF magnitude in HSC.
 The blue points show the original fraction in our sample, while the orange points present the completeness after removing the contamination of galaxies and including the incompleteness from the HSC survey.
 The errorbars are Poisson errors based on the numbers of corresponding objects. 
 }
\label{fig:comp}
\end{center}
\end{figure}

The second is the magnitude limits, which are determined by the surveys and the objects discarded during the matching algorithm.
As discussed in Section~\ref{ssec:S82}, the separation of SDSS objects becomes less reliable as the objects become fainter. 
Therefore, we can only trust the numbers in HSC for stars.
However, HSC data misses a significant number of bright objects, which are not of our primary interest, though the S82 observations are reliable at the bright end. 
The HSC star sample becomes complete with an $i$-band magnitude fainter than 19.
To calculate the completeness of our matched sample, we may combine the bright part of the S82 catalogue and the median and faint part of the HSC catalogue. 
Fortunately, our interests in this work lie in the faint WDs.  
Therefore, to simplify the completeness estimation, we adopt the sample with $i$-band PSF magnitude in HSC from 19 to 24.
Besides, we consider the completeness and contamination from the HSC survey following \cite{2018PASJ...70S...5B} and \cite{2018PASJ...70S...8A}. 
Considering the typical seeing of $0.6^{\prime\prime}$, the separation of star/galaxy in HSC at a magnitude of $i=24$ becomes less reliable, while the completeness of stars is down at about 70\%. 
The contamination of galaxies, which were classified as stars in HSC, is about 25\%.
The completeness of the selected stars is presented in Figure~\ref{fig:comp}. 
The blue points in the figure represent the fraction of matched objects of the HSC-identified stars. 
To ensure accurate measurements and eliminate contamination from HSC galaxies, we made a slight modification to recover the incompleteness from the survey. 
This modification is based on Figure 13 in \cite{2018PASJ...70S...8A}, as shown below,
\begin{align}
    P_{\mathrm{comp}}=P_{\mathrm{comp,match}}\times\frac{P_{\mathrm{{comp,HSC}}}}{
    (1-P_{\mathrm{{cont,HSC}}})}
\end{align}
where $P_{\mathrm{comp}}$ represents the final completeness, while $P_{\mathrm{comp,match}}$ is the completeness in our matching algorithm. $P_{\mathrm{comp,HSC}}$ and $P_{\mathrm{cont,HSC}}$ are the completeness and contamination from the HSC survey, respectively.
By recovering the completeness as a function of magnitude, we can obtain the expected WD densities. 
However, we want to again stress that, as long as we focus on large proper-motion objects such as WDs, the contamination of galaxies is smaller and unlikely to affect the main results of this paper. 
%

%%%%%%%%%%%%%%%%%%%%%%%%%%%%%%%%%%%%%%%%%%%%%%%%%%
\section{Selection Criteria of WD candidates}
\label{sec:selection}
The WD locus is distinctly positioned in the Hertzsprung-Russell (HR) diagram, diverging from main-sequence stars and other celestial objects, attributable to the low luminosity inherent to WDs. 
A notable magnitude difference of approximately ten exists between WDs and main-sequence stars sharing the same colour.
However, the absence of parallax in our catalogue impedes the derivation of absolute magnitude for our objects. 
Consequently, we try to apply the reduced proper motion (RPM) as a substitute for absolute magnitude to distinguish between WDs and other objects \citep[also see][for the similar idea]{2022MNRAS.510..611T}. 
This method has been empirically validated as an effective approach to obtaining a clean sample of WDs (e.g., \citetalias{2006AJ....131..571H}).

Meanwhile, the unprecedented photometric data provided by HSC presents a valuable opportunity to fit with the WD spectrum models for our prospective WD candidates.
Furthermore, the application of blackbody fitting emerges as an efficient strategy to identify promising WD candidates. 
The spectral energy distributions (SEDs) of WDs, especially featureless ones, such as DB WDs with a temperature too low to develop helium absorption lines or DC WDs with continuous spectrum, closely resembling blackbody spectrum \citep{2018AJ....156..219S}.
We will adopt both methods to identify the WD candidates and apply the corresponding weights that are used when estimating the luminosity function of WDs.

\subsection{Selection with reduced proper motion}
\label{ssec:RPM}
\subsubsection{Fiducial cut of reduced proper motion}
The relationship between distance and proper motion for celestial objects reveals a direct correlation: objects that are closer to us generally exhibit larger proper motions and vice versa. 
This relationship finds expression in the concept of reduced proper motion (RPM) \citep{1972ApJ...173..671J}, represented as $H_m$, 
which integrates the proper motion and apparent magnitude to facilitate an approximate estimation of absolute magnitude,
\begin{align}
H_m&=m+5\log\frac{\mu}{1~{\rm arcsec}~{\rm yr}^{-1}}+5\nonumber\\
&=M+5\log v_\mathrm{t}-3.38
\end{align}
where $\mu$ is the proper motion in units of arcseconds per year, 
$m$ is the apparent magnitude, 
$M$ is the absolute magnitude, 
and $v_\mathrm{t}$ is the tangential velocity in units of kilometres per second. 
For a specific colour, the distribution of RPM represents the combination of the absolute magnitude and the tangential velocity.
The implicit assumption of RPM is that if faint stars in the solar neighbourhood follow a population with similar peculiar velocity $(v_\mathrm{t}$),
the measured proper motion and apparent magnitude give an estimate of the absolute magnitude.
The deep HSC data enables us to probe WDs and stars in the thick disc and halo regions, 
which tend to have larger velocities with respect to the Sun (see below).
Such WDs give an even larger RPM than thin-disc WDs do at a given colour. On the other hand,
stars in the thick disc and halo could give contaminations to the region of relatively bright WDs in the RPM diagram.

\begin{figure}
\begin{center}
 \includegraphics[width=\columnwidth]{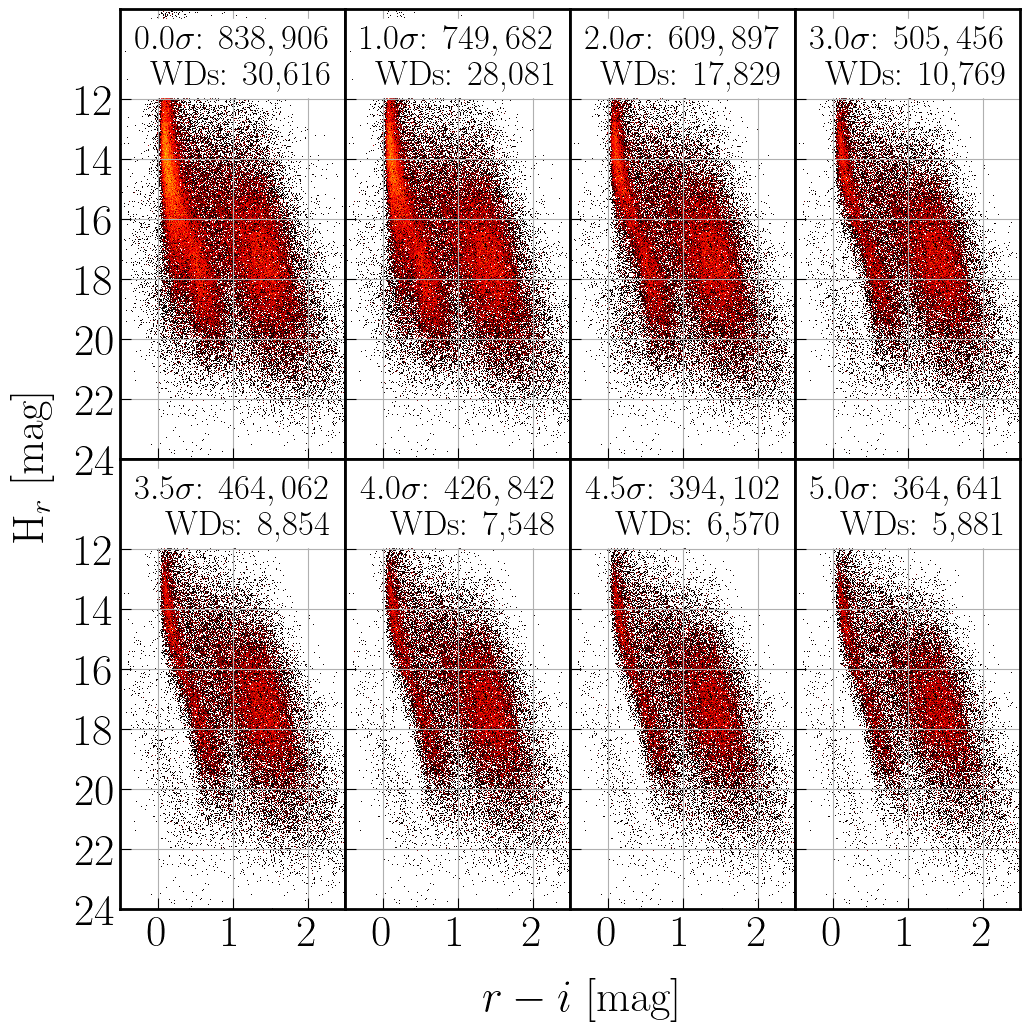}
 \caption{The reduced proper motion diagrams of objects with different cuts of the detection signal-to-noise ratio ($\mu/\sigma_\mu$)}
 ranging from 0 to 5. 
 The warmer colours indicate higher density, while individual objects are depicted as black dots.
 The numbers in each panel denote the numbers of total star objects and WDs in our catalogue, respectively.
 We can find a clearer separation between stars and WD locus as the significance increases. 
 The other two sequences right to the WD locus are contributed by the stars belonging to the halo and the disc, respectively.
 
 \label{fig:8_RPM}
\end{center}
\end{figure}

WDs are intrinsically much fainter than main-sequence (MS) stars (by about 10 magnitudes), while their velocities are comparable.
Therefore, there is a gap between WDs and other objects (MS stars, dwarfs, giants, etc), 
which gives an efficient way to separate WDs and other objects to obtain a pure WD sample. 
As shown in Figure~\ref{fig:8_RPM}, we present the RPM diagrams for different cuts of the proper motion detection 
signal-to-noise ratio ranging from $\mu/\sigma_\mu=0$ to $5$.
Here we calculate the RPM, ${\rm H}_r$ for each object using the HSC $r$-band magnitude and the proper motion (Section~\ref{ssec:PM}). 
As the proper motion significance increases, the separation is more evident.
We choose the 4$\sigma_\mu$ threshold as our default threshold, which has removed most non-moving objects that are unlikely to be WDs. 

\begin{figure}
\begin{center}
 \includegraphics[width=\columnwidth]{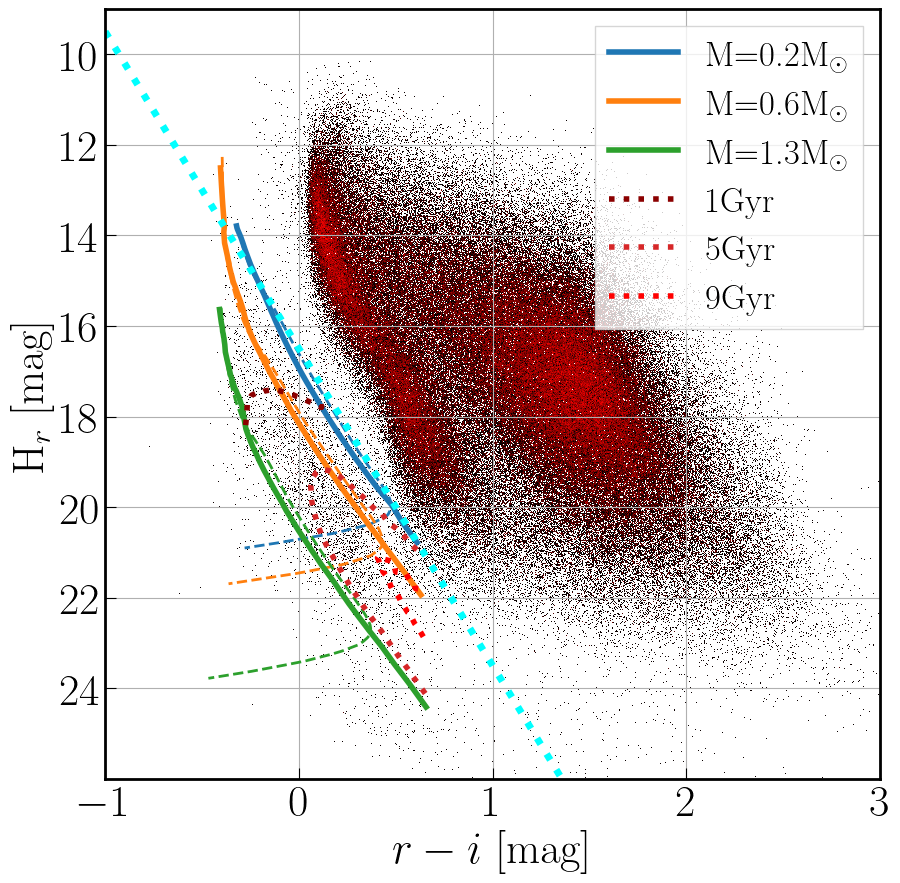}
 \caption{The reduced proper motion diagram of objects selected by the $\mu/\sigma_\mu$} cut of proper motion 
 measurements: $\mu>4\sigma_\mu$.
 The solid lines represent the cooling tracks of the He atmosphere model of WD (assuming $v_\mathrm{t}=40$~km~s$^{-1}$) with different masses as shown in the legend, respectively, while the dashed line shows the cooling sequence of H atmosphere WD model.
 The red dotted lines show the corresponding ages of the cooling sequence with the tangential velocity $v_\mathrm{t}=40$~km~s$^{-1}$ and mass $M=0.6M_\odot$.
 The dotted cyan line presents our threshold shown by Eq.~(\ref{equ:2}), which can nearly include most WD candidates with the tangential velocity above our lower limit (40~km~s$^{-1}$).
 
 \label{fig:RPMmodel}
\end{center}
\end{figure}

Figure~\ref{fig:RPMmodel} shows a zoom-in distribution of objects selected with the detection threshold of $\mu/\sigma_\mu>4$.
We also show the cooling tracks assuming different WD masses of 0.2, 
0.6 and 1.3 solar masses, respectively.
The solid lines show the He-atmosphere model, while the dashed line shows the H-atmosphere model.
We derive the cooling tracks using the \texttt{PYTHON} package \texttt{WD\_models}\footnote{\url{https://github.com/SihaoCheng/WD_models}} developed by Sihao~Cheng, which are based on the model from \cite{2020ApJ...901...93B}\footnote{\url{http://www.astro.umontreal.ca/~bergeron/CoolingModels/}}.
This package is partly similar to the \texttt{WDPhotTools} mentioned above, but more simple and flexible to make the cooling plots. 
The relevant model of WDs corresponds to the work by \cite{2011ApJ...730..128T} and \cite{2011ApJ...737...28B}, while the details of the colour calculations are given in \cite{2006AJ....132.1221H}, which represent an extension of the earlier work by \cite{1995PASP..107.1047B}.

WDs typically exhibit a mass distribution that peaks around 0.6$M_\odot$ \citep{2007MNRAS.375.1315K}, with variations observed based on their spectra,
while the surface gravity tends to have a narrow distribution around $\langle\mathrm{log}\ g\rangle=7.998\pm 0.011$, as reported by \cite{2015MNRAS.446.4078K,2016MNRAS.455.3413K,2021MNRAS.507.4646K}. 
Hence we empirically determine the following RPM cut to select WD candidates:
\begin{align}
    H_r > 7(r-i) +16.5
    \label{equ:2}
\end{align}
where $r$ and $i$ are the apparent magnitudes in HSC.
The coefficients of $7$ and 16.5 are chosen empirically.
As can be found from Figure~\ref{fig:RPMmodel}, the above cut can select extremely 
light WDs of $0.2M_\odot$ if those have $40$~km~s$^{-1}$. 
Since WDs could have a smaller tangential velocity, we will employ the cut of $v_{\rm t}>40$~km~s$^{-1}$
to include most WD candidates while avoiding contamination from subdwarfs having large velocities.
We will again discuss the impact of the velocity cuts on our results, using the standard Milky Way model of kinematic distribution for stars. 

\subsubsection{Measurement uncertainty of proper motion}
\label{sssec:lowerlimit}
We select our WD candidates from stars with at least $4\sigma_\mu$ detection of the proper motions. 
The main source of the proper motion measurement error stems from uncertainties in the centroid determinations of stars, as reported in the HSC and S82 catalogues. 
However, as we discuss in this section, these centroid errors tend to underestimate the actual measurement errors.
The measurement errors are the result of a combination of various factors in our proper motion measurements, including photometry, centroid determination, astrometry calibrations, spatial variations in data quality,
imperfect calibrations, and so on.
Therefore, we here estimate these errors directly from the data themselves.

Galaxies and quasars should have no proper motions if the measurement is perfect. Hence, 
we use quasars and galaxies to estimate the net measurement error, following the method in Appendix~C of \citetalias{2021MNRAS.501.5149Q}. 

\begin{figure}
\begin{center}
 \includegraphics[width=0.9\columnwidth]{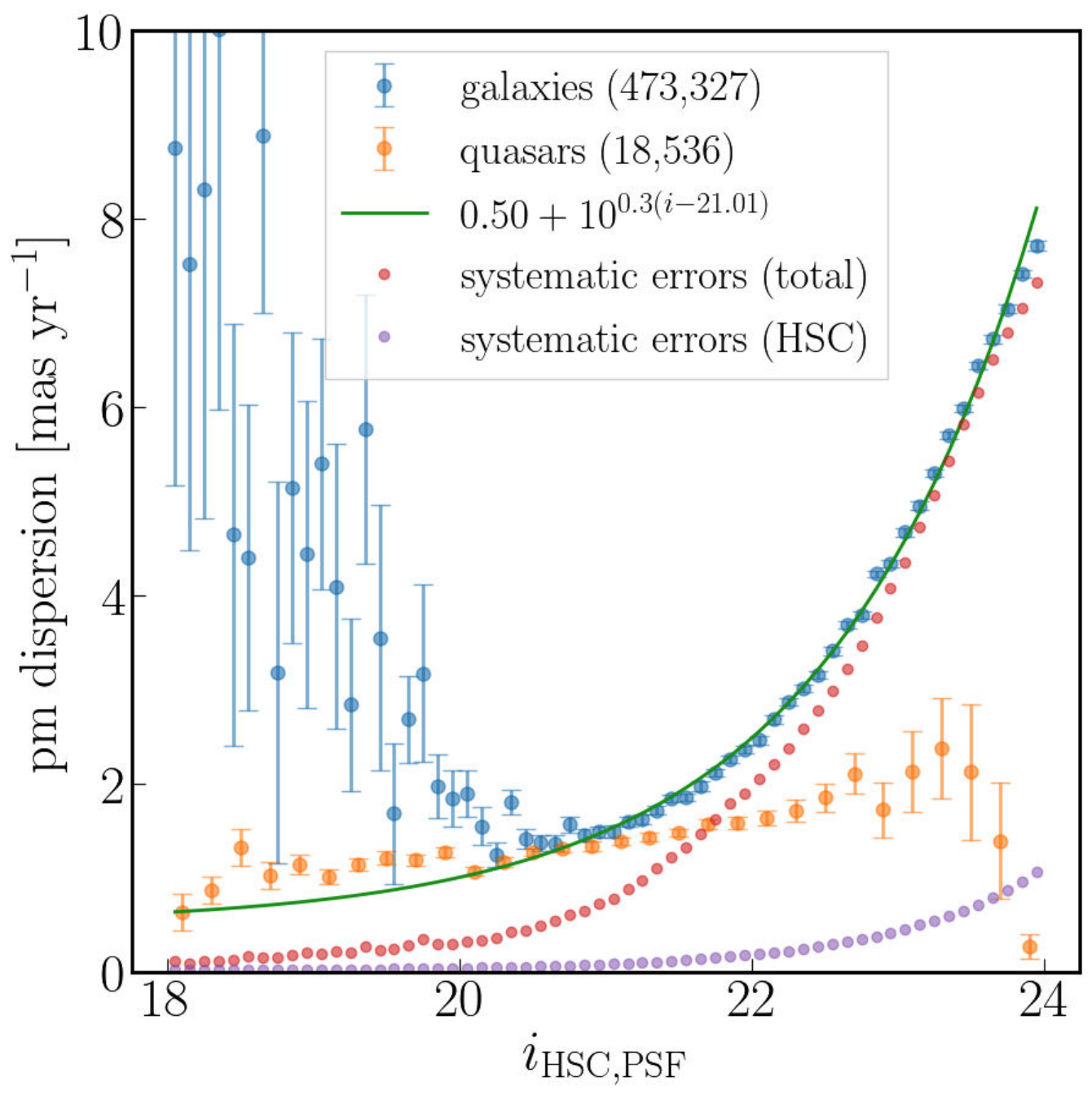}
 \caption{Statistical errors in our proper motion measurements.
 Blue points represent errors estimated from the dispersion or width of the apparent proper motion distribution of ``small-size'' galaxies in each bin of the HSC $i$-band PSF magnitudes.
 The orange points represent errors estimated from quasars, which are sourced from the SDSS DR14 quasar sample and matched to our catalogue.
 The green line illustrates the expected trend of the error with a specific flux of the object.
 The data points of galaxies align nicely with this expectation.
 The red points represent errors estimated from the quadrature sum of the centroid errors of galaxies in the S82 and HSC catalogues, while the purple points are derived from the centroid errors in the HSC catalogue alone.}
   
 \label{fig:pm_err}
\end{center}
\end{figure}

These proper motion errors are mainly dependent on magnitudes (the HSC $i$-band magnitudes in our case). 
As shown in Figure~\ref{fig:pm_err},
we divide the galaxies and quasars into different $i$-band magnitude bins, spaced by 0.1~mag, and then 
estimate the measurement error from the scatters of the ``apparent'' proper motions of the galaxies and quasars in each bin.
We then model the relationship between the measurement error and the $i$-band magnitude by the following function,
\begin{align}
    \sigma_\mu(i)=a+10^{0.3(i-i_0)}
    \label{eq3}
\end{align}
where $i$ is the magnitude in HSC PSF photometry, $i_0$ is a pivot magnitude, and a parameter $a$ is introduced to model a magnitude-independent error contribution.

We select small-size galaxies, which are defined from the HSC data based on the criterion $-0.3<i_{\rm HSC,cModel} - i_{\rm HSC, PSF}<-0.15$, to avoid the additional scatters from uncertainties in the centroid determination while measuring proper motions.
Quasars are also employed to represent the bright part of the measurement error, as the galaxies in HSC are insufficient to illustrate this aspect.
We have identified 473,327 small-size galaxies in the HSC-S82 regions, and the error bars are estimated using the bootstrap method in each magnitude bin. 
To avoid large scatter in the bright end, we only use the data with $i$-band magnitude ranging from 21 to 24. 
We have determined the coefficients for Eq.~\ref{eq3} as $a=0.50$ mas~yr$^{-1}$ and $i_0=21.01$ mag. 
Subsequently, we will employ it to estimate the proper motion errors in the following sections of this work.
For the brighter range, the estimated uncertainty is slightly lower than that derived from quasars. 
However, we do not consider this discrepancy to be problematic for two reasons.
Firstly, the selection of WDs via the reduced proper motion diagram will filter out objects with low proper motion from the WD candidate pool. 
More importantly, the observed bright WDs are expected to be in close proximity to us, which will influence the limits of the discovery fraction integration, as described in Section~\ref{ssec:DF}. 
The lower tangential velocity limit of 40~km\,s$^{-1}$ is significantly larger than the limit determined from proper motion uncertainty.

\subsubsection{Completeness: the uppercut of proper motion}
\label{sssec:upperlimit}
The number counts of stars against proper motions is useful to infer the completeness of our sample. 
For reference, we consider a case that stars follow a uniform number density that would be a good assumption for stars in the solar neighbourhood.
In this case, the number counts of stars in the distance range of $[d,d+\mathrm{d}d]$ is given as 
\begin{align}
    \frac{\mathrm{d}N}{\mathrm{d}d}=\rho \Omega_A d^2\propto d^2
\end{align}
where $\rho$ is the number density profile for which we assume is constant $(\rho\propto d^0)$ and $\Omega_A$ is the solid angle of the survey area ($\sim 165$~deg$^2$ for our S82 case). Hence, the cumulative number count is 
\begin{align}
    N(>\mu)&\propto d^3 \propto v_\mathrm{t}^3\mu^{-3}
\end{align}
where we have used $v_\mathrm{t}\propto d\times \mu$ for the relation between the tangential velocity and the proper motion. Hence if stars follow a constant velocity, we expect that the cumulative number count scales with $\mu$ as $N(>\mu)\propto \mu^{-3}$. 
\begin{figure}
\begin{center}
 \includegraphics[width=0.9\columnwidth]{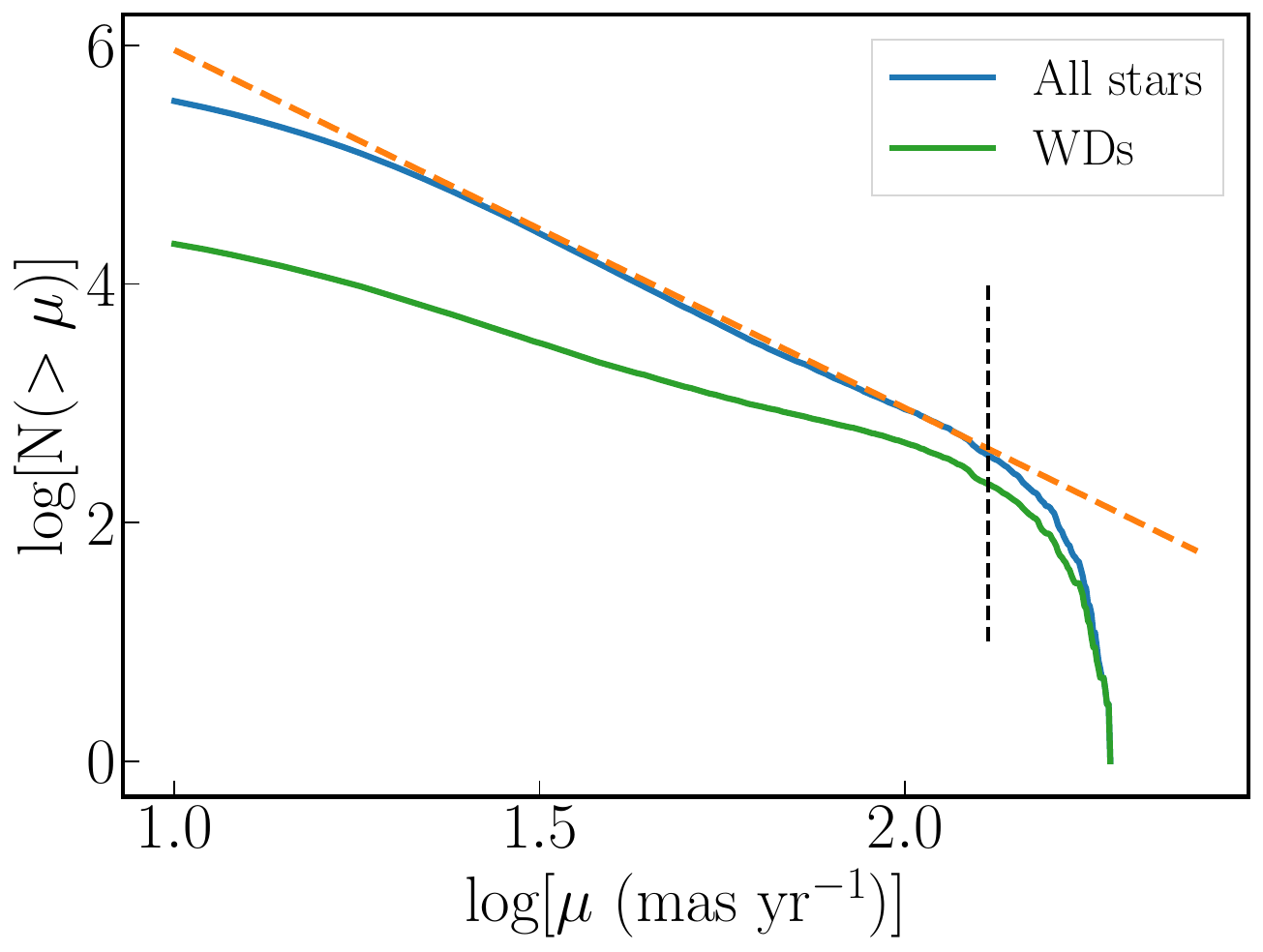}
 \caption{The cumulative number counts all stars and WDs with proper motion ($\mu$) larger than a given value plotted in the $x$-axis.
 For comparison, the orange-dashed line shows the counts of $N(>\mu)\propto \mu^{-3}$ that is an expected behaviour for objects with uniform number density and constant tangential velocity (see text for details).
 The Black dashed line represents a proper motion cut at 130~mas~yr$^{-1}$ which we impose for our analysis.
}
 \label{fig:comp_pm}
\end{center}
\end{figure}
Figure~\ref{fig:comp_pm} shows the cumulative number counts for stars (including WDs) and WD candidates, where WDs are selected based on the RPM cuts in Figure~\ref{fig:RPMmodel}. It is clear that the star sample very closely follows a power-law of $\mu^{-3}$ over the wide range of $\mu$, indicating that stars with large proper motions at $\mu\gtrsim 30~{\rm mas}~{\rm yr}^{-1}$ are dominated by stars in the solar neighbourhood.
The counts start to deviate from the power law at $\mu\simeq 130~{\rm mas}~{\rm yr}^{-1}$,
indicating an incompleteness in our matching process beyond this proper motion threshold. 
This deviation is primarily attributed to the searching radius of $2^{\prime\prime}$ employed in our matching algorithm.
Additionally, the deviation in the low proper motion region is a consequence of the contribution of distant stars outside the solar neighbourhood and/or 
the inaccuracy (larger errors) of proper motion measurement.
On the other hand, the WD sample shows a shallower power-law slope and even exhibits a break in the power-law index at $\mu\sim 60~{\rm mas}~{\rm yr}^{-1}$. 
The WDs have a more substantial contribution to objects with greater proper motion.
This is attributed to the fact that faint WDs with lower proper motion, as well as bright MS stars with higher proper motion, are likely to be excluded due to magnitude limits.
The WD sample also indicates an incompleteness at 
$\mu\simeq 130~{\rm mas}~{\rm yr}^{-1}$.
Therefore, in this study, we have chosen $\mu=130~{\rm mas}~{\rm yr}^{-1}$ as the uppercut of $\mu$ in 
our sample: that is, 
we use objects with $\mu<130~{\rm mas}~{\rm yr}^{-1}$ for the following results.

\subsection{Temperature estimate of each WD candidate}
\label{sec:spectral_type_estimation}
To further refine the screening of WD candidates, we fit a WD atmosphere model to the multi-band HSC photometries of each WD candidate. 
There are 5 filters in the HSC system whose effective central wavelengths are
$(g,r,i,z,y)=(4754,6175,7711,8898,9762)$\AA~ \citep[see Table~3 in][]{2018PASJ...70S...4A} \citep[also see][]{2018PASJ...70S...1M,2018PASJ...70...66K}, respectively. 
This fitting also enables us to estimate the absolute magnitude or equivalently distance from the observed apparent magnitude (flux) for each WD candidate by assuming a typical radius of WD.
The estimation of absolute magnitude is crucial for estimating the WD luminosity functions.
In this paper we use the following two methods to estimate an atmosphere temperature for each WD: i) a blackbody model and ii) the template WD atmosphere model. In the following, we describe these two methods.

\subsubsection{Blackbody fitting}
\label{ssec:selectBBF}
The blackbody fitting method is simple.
The HSC photometries 
cover the turnover of the blackbody spectrum for objects with relatively low temperatures ($T\lesssim6000$ K).
The blackbody fitting gives a robust way to validate the WD candidates.
The blackbody flux follows
\begin{align}
f_\lambda=a\frac{2hc^2}{\lambda^5}\frac{1}{\mathrm{exp}(hc/\lambda k_\mathrm{B} T_\mathrm{eff})-1}
\label{eq:blackbody_spectrum}
\end{align}
where $\lambda$ is the wavelength, $h$ is the Planck constant and $k_\mathrm{B}$ is the Boltzmann constant. 
The fitting parameter $T_\mathrm{eff}$ is the effective temperature and $a$ is the normalisation factor (solid angle), which is given by the distance to the object, $d$, and the radius of the object, $R$, as $a=\pi R^2/d^2$.
We apply the blackbody spectrum fitting to each of the WD candidates to estimate the effective temperature and the normalisation factor. 
It is important to note the photometric uncertainty in the catalogue, which accounts only for statistical errors from the model fitting of the PSF photometry.
The systematic error is estimated to be around 1\% \citep{2018PASJ...70S...6H,2018PASJ...70S...8A,2022PASJ...74..247A}, and we take this into account in the fitting by combining it with the statistical error using the formula $\sigma_\mathrm{tot}=\sqrt{\sigma_\mathrm{sys}^2+\sigma_\mathrm{stat}^2}$.
The systematic error mainly affects the photometric fitting of bright objects in the HSC data, where the statistical error provided in the catalogue is extremely small, less than $0.001$~mag for objects with $i$-band magnitude brighter than 20~mag. 
For faint objects with $i>23$ mag, the typical uncertainties in the $grizy$ bands are 0.04, 0.03, 0.02, 0.05, and 0.11 mag, respectively.
When converting from flux to magnitude using the simple error propagation formula $|\Delta m|=\frac{2.5}{\mathrm{ln} 10}(\frac{\Delta f}{f})$, we find that the systematic error does not significantly impact our analysis of the faint  WD candidates.
We collect the reduced chi-square $\chi^2_\nu$ to evaluate the goodness-of-fit with the degree of freedom $\nu=3$.
There are 2,393 objects with good fitting results ($\chi^2_\nu<2$) in total and 1,790 objects 
in the temperature range from 6,000 to 10,000~K.
\begin{figure}
\begin{center}
\includegraphics[width=\columnwidth]{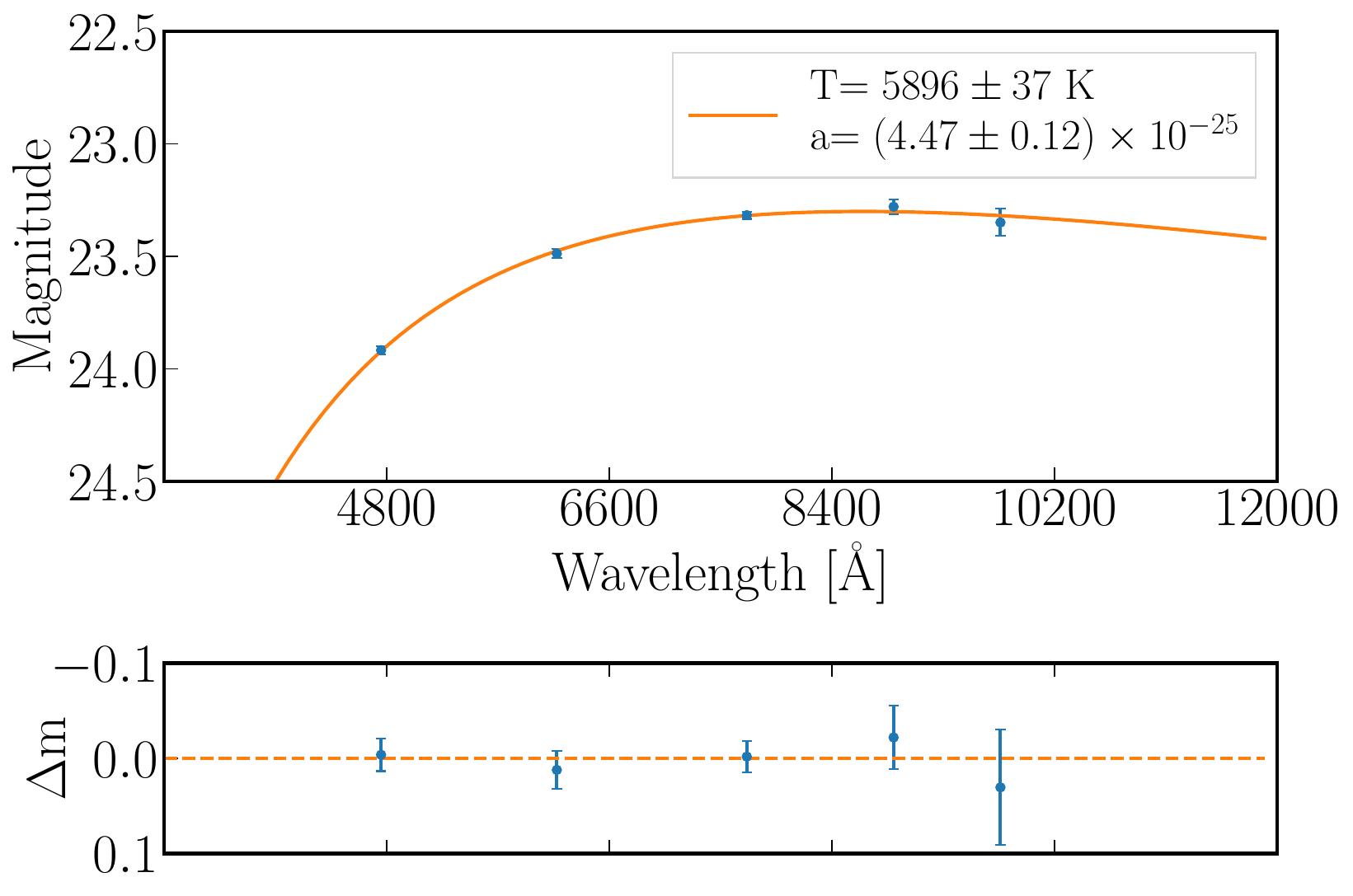}
 \caption{One example of blackbody spectrum fitting to a WD candidate. 
 At the bottom panel, we also show the magnitude residuals from the fit.
 The goodness-of-fit is 0.37 in this sample. 
  }
 \label{fig:bbf}
\end{center}
\end{figure}
In Figure~\ref{fig:bbf}, we give an example of a WD candidate that has a good blackbody fit.
The legend displays the estimated values of the temperature and the normalisation factor. 

\subsubsection{Fitting with WD atmosphere models}
\label{ssec:WDmodel}
In addition to the blackbody fitting method, we use a more sophisticated model of WD atmosphere model to fit the photometries of each WD candidate.
There has been significant progress in studying the atmospheres of WDs, thanks to the availability of abundant spectroscopic data from surveys like SDSS \citep[e.g.,][]{2004ApJ...607..426K, 2013ApJS..204....5K, 2016MNRAS.455.3413K, 2021MNRAS.507.4646K}.
By assuming a surface gravity, we can fit the photometric data of a WD sample with different atmosphere models of WD.
This approach has been demonstrated in various studies that compare photometric information with spectroscopic solutions to calibrate the atmosphere models \citep[e.g.,][]{2014ApJ...796..128G,2019ApJ...871..169G,2019ApJ...882..106G}.

In this paper, we use the WD atmosphere model given by
the public software \texttt{WDPhotTools}\footnote{\url{https://github.com/cylammarco/WDPhotTools}} in \cite{2022RASTI...1...81L}.
The fitting process involves applying pure H and He atmosphere models to a given photometric data. 
Since the number of other types of WDs is relatively rare, our focus is on fitting with H and He models.

The steps involved in the fitting process are as follows:
\begin{enumerate}
    \item We employ the Markov chain Monte Carlo (MCMC) sampling technique to explore the parameter space and generate initial estimates for the temperature, surface gravity, and distance for both the H and He atmosphere models. 
    The \texttt{emcee}\footnote{\url{https://github.com/dfm/emcee}}\citep{2013PASP..125..306F} package is used in this progress.
    \item We set the surface gravity log $g$ as a known parameter and treat the temperature and distance as the initial guesses. 
    We aim to minimise the $\chi^2$ statistic, which quantifies the difference between the observed photometric data and the model predictions. 
    To achieve this, the code utilises the \texttt{scipy}\footnote{\url{https://scipy.org}} package and employs optimisation algorithms to find the best-fit values for temperature, distance, and subsequently, the bolometric magnitude. 
    \item By following this procedure, we can obtain the best-fitting results for the properties of each of our WD candidates with H and He models, respectively.
\end{enumerate}
The choice between H and He atmosphere models for WDs has a limited impact on the derived luminosity for temperatures above approximately 6,000 K. 
At these higher temperatures, assuming an H atmosphere for a He atmosphere WD will result in a minor overestimation of the luminosity and a slight underestimation of the distance.
However, below 6,000~K, using an incorrect atmosphere type can lead to significant errors in the derived luminosity and distance. 

The choice of the upper reduced $\chi^2$ cut has been set at $\chi_\nu^2=10$ for WDs with temperatures above 6,000~K and $\chi_\nu^2=2$ for WDs below 5,000~K. Between these temperature ranges, a linear transformation is applied to transition between the two thresholds.
When the temperature is too high, the tail of the spectrum becomes less sensitive to changes in temperature, which means that photometric fitting can be imprecise.
By setting a higher maximum reduced $\chi^2$ value, we allow for a broader tolerance in the fitting process, as the deviations from the model are expected to be less significant due to the spectral characteristics at higher temperatures.
On the other hand, for low temperatures, where the sensitivity of the spectrum to temperature changes is higher, a lower maximum reduced $\chi^2$ value is applied to ensure a more stringent fit, minimising the potential deviations and uncertainties in the modelling.
Besides, the low-temperature objects are of particular interest in our study as they represent a novel sample that provides valuable insights into the oldest WDs in our Galaxy.

\subsubsection{Weights}
\label{ssec:weights}
As we described above (Sections~\ref{ssec:selectBBF} and \ref{ssec:WDmodel}), we performed a model fitting for 
each of our WD candidates with both the blackbody and the WD atmosphere model. We keep only the WD candidates that pass the $\chi^2_\nu$ cuts as an indicator of the goodness-of-fit. 
To obtain the intrinsic properties of each WD candidate, the effective temperature and the solid angle $a(=\pi R^2/d^2)$, we take into account the fitting results from the blackbody and the H and He WD models by 
using the weights given by their respective likelihoods,
$P_\mathrm{BB}\propto\mathrm{exp}(-\chi^2_\mathrm{BB}/2)$, 
$P_\mathrm{H}\propto\mathrm{exp}(-\chi^2_\mathrm{H}/2)$, 
and $P_\mathrm{He}\propto\mathrm{exp}(-\chi^2_\mathrm{He}/2)$, respectively.

\begin{figure}
\begin{center}
\includegraphics[width=\columnwidth]{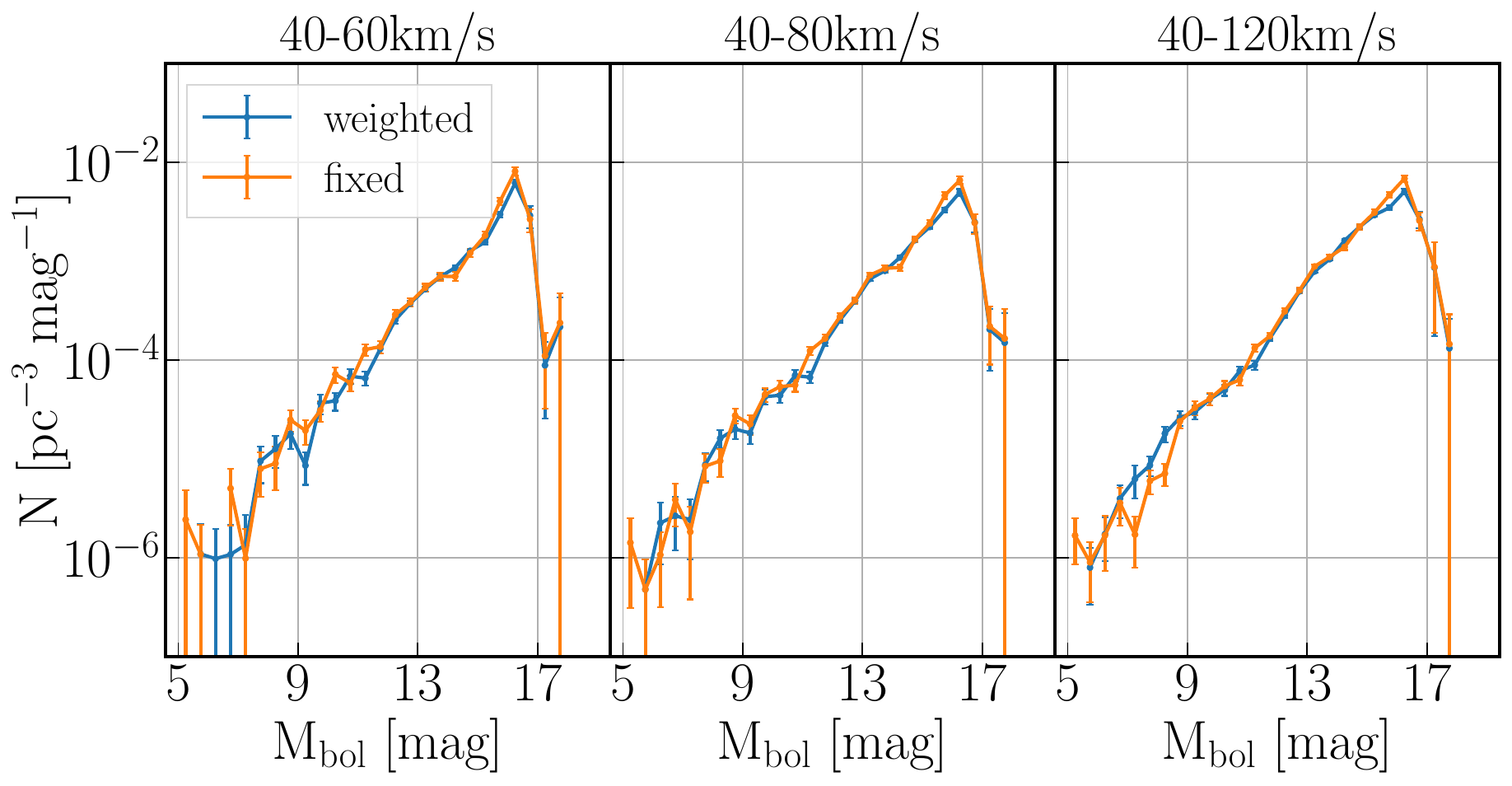}
 \caption{Comparison of the luminosity function results derived from the weighted fitting model and the fixed model. 
 The three panels correspond to samples with different tangential velocity ranges, as we will explain the details in Section~\ref{sec:WDLF}. 
 The results are similar for both methods.
 }
 \label{fig:weight}
\end{center}
\end{figure}

By adopting weighting, we aim to obtain a more precise estimation for candidates that resemble either a blackbody or a pure H/He model with comparable reduced chi-squared values ($ \chi^2_\nu $). 
For most candidates, however, only one model is dominant. 
Consequently, the differences between the fixed model and the weighted model are quite small and can be neglected. 
We also attempted to fix the spectral type for each object based on their $ \chi^2_\nu $. Figure~\ref{fig:weight} shows a comparison of the luminosity functions derived from the weighted fitting model and the fixed model across three different velocity ranges. 
The luminosity function results are similar between the two methods.
We will introduce the details of the luminosity function in the following Section~\ref{sec:WDLF}.
\begin{table}
 \begin{center}
 \begin{tabular}{lcr}
  \hline
  \hline 
  Selection & Description & Number\\[2pt]
  \hline
  Total WD candidates & $H_r > (r-i)\times 7 +16.5$ & 30,616\\[2pt]
  Proper motion  & $\mu>4\sigma_\mu(i)$ & 7,548\\[2pt]
  Magnitude & $19<i<24$ & 7,305\\[2pt]
  Fitting & $\chi^2_\nu<10$ & 5,803\\[2pt]
  Final & $v_\mathrm{t}>40$ km s$^{-1}$ & 5,080\\[2pt]
  \hline
 \end{tabular}
 \end{center}
 \caption{The table provides the number of WD candidates that passed the selection criteria. 
 The first row corresponds to the total number of WD candidates that have a reduced proper motion larger than our threshold based on the $r-i$ colour (Figure~\ref{fig:RPMmodel}). 
 The second row represents the lower limit for proper motion, which is set at $4\sigma_\mu$ as a function of the $i$-band magnitude. 
 The third row denotes the magnitude limits in the $i$-band which are determined based on considerations of completeness and contamination. 
 The fourth row denotes the maximum tolerance for the goodness-of-fit that is given by the reduced chi-square, denoted as $\chi^2_\nu$. 
 The last row indicates the number passed the strict selection of the fitting models and the tangential velocity limit, which will be taken into account as a weight in the subsequent calculations.
}
 \label{tab:WDnum}
\end{table}
\subsection{Photometric distance}
\label{ssec:VmaxDist}
As described in Section~\ref{sec:spectral_type_estimation},
we estimate the intrinsic properties for each of the WD candidates from the model fitting. 
The estimated solid angle parameter, $a$ (Eq.~\ref{eq:blackbody_spectrum}), is determined by the radius of the WD and its distance from us, given by $a=\pi R^2/d^2$. 
For WDs, the surface gravity distribution typically peaks around log $g = 8.0$, as shown in the SDSS DR16 work \citep{2021MNRAS.507.4646K}, with a mean mass of about 0.6$M_\odot$. 
By assuming a radius of approximately $0.013R_\odot$, we can estimate the distance from the best-fit $a$ parameter for each WD candidate.
The assumption of a constant radius for WDs is based on the narrow distribution of their masses. 
The 15\% mass dispersion covers about 70\% of WDs, indicating that most WDs have relatively similar masses. 
Considering the resulting error from the 1$\sigma$ mass dispersion, which represents $0.1 M_\odot$, the uncertainty in the photometric distance would be around 5\%.

It is important to note that the photometric distance estimation for WDs could still contain a residual uncertainty. 
This is mainly due to the limited study of WDs and the challenges associated with their characterisation.
Additionally, our fitting process is based on the HSC 5-band photometries, which may not provide sufficient information to accurately determine the spectral energy distribution, particularly for faint objects.
Therefore, the photometric distance estimates should be interpreted with caution and further studies are needed to improve their accuracy.

\begin{figure}
\begin{center}
 \includegraphics[width=\columnwidth]{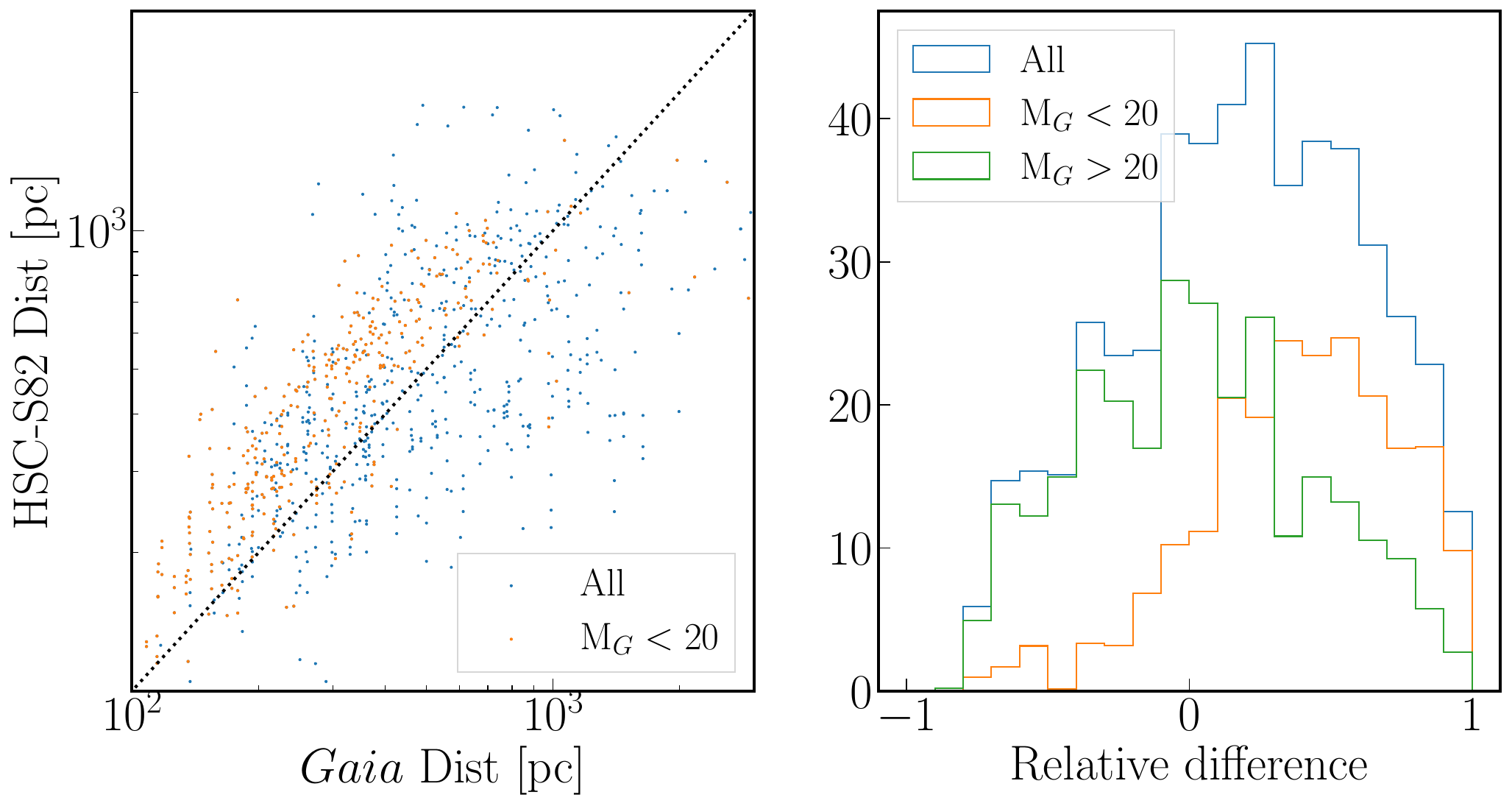}

 \caption{Comparison of the photometric distances from the HSC-S82 catalogue with the parallax distances from {\it Gaia}. 
 In the left panel, we show the photometric distances from HSC-S82 versus the parallax distances from {\it Gaia}.
 The right panel presents a histogram of the relative differences in the distance ($\Delta r = (r_{\mathrm{HSC-S82}} - r_{\it Gaia}) / r_{\it Gaia}$). 
 The photometric distance results tend to overestimate compared to those from {\it Gaia} for brighter WDs ($G < 20$ mag), while the fainter WDs ($G > 20$ mag) show better consistency, although they both suffer from larger uncertainties due to the faintness.
 }
\label{fig:Gaia_dist}
\end{center}
\end{figure}

We also compare the photometric distances from HSC-S82 with the parallax distances from {\it Gaia}. 
As mentioned in Section~\ref{ssec:comparisonwithgaia}, there are 546 WDs matched between HSC-S82 and {\it Gaia}. 
As shown in Figure~\ref{fig:Gaia_dist}, the brighter objects ($G < 20$ mag) show a more significant difference between our estimates and those from {\it Gaia}, while the fainter objects ($G > 20$ mag) are more consistent, although there is still considerable scatter.
Indeed, most WD candidates in this work are fainter than the magnitude limit of {\it Gaia}.
Therefore, we would expect better consistency with the parallax distances from {\it Gaia}. 
However, the distance estimations from {\it Gaia} also suffer from large uncertainties for these distant objects. 
In the subsequent analysis of the luminosity function, if our method has indeed overestimated the distances, then the derived absolute magnitudes would be underestimated (i.e., appear brighter), which would shift the luminosity function towards the bright end. 
Therefore, any potential overestimation would only weaken our conclusions; however, we still obtain a large sample of faint WDs.

Once the distance of each WD candidate is estimated, we can estimate the tangential velocity from the measured proper motion: 
\begin{align}
v_t~[{\rm km}~{\rm s}^{-1}]\simeq 4.74\times d~[{\rm kpc}]\times \mu~[{\rm mas}~{\rm yr}^{-1}].
\label{eq:vt_def}
\end{align}
For instance, a WD with $\mu=100~{\rm mas}~{\rm yr}^{-1}$ and $d=1~{\rm kpc}$ has $v_{\rm t}\simeq 474~{\rm km}~{\rm s}^{-1}$.

\subsection{Summary of our WD candidate sample}
\label{ssec:summary_wd_sample}
Table~\ref{tab:WDnum} summarises the number of WD candidates that passed our selection cuts. 
Our cuts are, in sequential order, based on the reduced proper motions (Section~\ref{ssec:RPM}), the fitting result of the HSC photometries with the WD atmosphere models (Section~\ref{sec:spectral_type_estimation}), and then the tangential velocity. The rationale to adopt the tangential velocity cut of $v_{\rm t}>40~{\rm km}~{\rm s}^{-1}$ 
is as follows. If WDs have an intrinsic small tangential velocity, those tend to be distributed in a similar region to that of faint stars (e.g. see Figure~4 of \citetalias{2017AJ....153...10M}). 
Such WDs with small tangential velocities are likely in the solar neighbourhood, i.e. the Galactic disc. Our primary focus is on distant WDs in the thick disc and halo regions, which can be uniquely studied by utilising deep HSC data. 
Hence, in order to remove the contamination of faint main sequence stars that have similar reduced proper motions to those of WDs, we adopt the cut of $v_{\rm t}>40~{\rm km}~{\rm s}^{-1}$.
In total, we use 5,080 WD candidates as the parent sample of our study.

%%%%%%%%%%%%%%%%%%%%%%%%%%%%%%%%%%%%%%%%%%%%%%%%%%
\section{Methodology}
\label{sec:method}
In this section, we describe our method to estimate the luminosity function of WDs.
We use the maximum volume density estimator \cite{1976ApJ...207..700F} for reconstructing the luminosity function, 
where we correct for the completeness effect or selection function of our WD sample following the method in \citetalias{2006AJ....131..571H}.

\subsection{Mock catalogue of WDs in the S82 region}
\label{ssec:mock}
To evaluate the selection function of our WD candidates, we construct a mock catalogue of WDs in the S82 region, using the standard model of the Milky Way. 
Here the implicit assumption we employ is that WDs follow the same spatial and kinematic distributions as those of stars, which should be a good assumption as WDs are remnants of main-sequence stars. 
In this subsection, we describe our mock catalogue.

\subsubsection{Density profile}
\label{ssec:density}
Our magnitude limit is $i\simeq 24$, which allows us to sample WD candidates at kilo-parsec distances.
Hence our WD candidates would belong to all three distinct regions of the Milky Way Galaxy, i.e. the thin and thick disc regions and the halo region.
To take into account these contributions, we use the Galactic model in \cite{2008ApJ...673..864J} to construct a mock catalogue of stars (equivalently WDs) in the three regions.

For the disc stars, we model their distribution as an exponential profile with scale-height $H$ and radial scale-length $L$ shown below,
\begin{align}
\rho_{\mathrm{disc}}(R,z+z_\odot)=\rho_{\mathrm{disc}}(R_\odot,0) \mathrm{exp}\left(\frac{R_\odot-R}{L}-\frac{|z+z_\odot|}{H}\right)
\end{align}
where $z$ is the vertical distance from the plane of our Sun which can also be represented as $d\mathrm{sin} b$ where $d$ is the distance to a star from us, and $b$ is the Galactic latitude. 
$R$ is the radial distance from the Galactic centre.
$\rho_{\mathrm{disc}}(R_\odot,0)$ is the maximum stellar number density for the disc at the solar radius $R_\odot$, on the Galactic plane with $z=-z_\odot$.
Given our line-of-sight direction, which is approximately 60 degrees from the Galactic plane, the density variations along the Galactic radius direction are relatively small compared to those along the vertical direction.
Consequently, we have simplified our approach by modelling the density profile as a function solely of the vertical distance from the plane of our Sun,
\begin{align}
\rho_{\mathrm{disc}}(z)=\rho_{\mathrm{disc}}(R_\odot,0)\mathrm{exp}\left(-\frac{|z+z_\odot|}{H}\right)
\end{align}
This approximation allows us to accelerate the generation of the mock catalogue. 
The solar distance from the Galactic plane is $z_\odot \sim 20$~pc. 
We further decompose the disc density profile into a combination of the sum of thin and thick discs with individual scale heights.
In this work, we set $H_\mathrm{thin}\sim 250$~pc for the thin disc and $H_\mathrm{thick}\sim 1,000$~pc for the thick disc, respectively. 
It is worth noting that the specific values of scale height can vary among different studies. 
However, the choice of scale height does not have a significant impact on the velocity distribution of the disc, which is the key factor in determining the discovery fraction for our calculation of the luminosity function.
Therefore, we have a combined density profile shown below,
\begin{align}
\rho_{\mathrm{disc}}(z)=\rho_{\mathrm{thin},0} \mathrm{exp}\left(-\frac{|z+z_\odot|}{H_{\mathrm{thin}}}\right)+\rho_{\mathrm{thick},0} \mathrm{exp}\left(-\frac{|z+z_\odot|}{H_{\mathrm{thick}}}\right)
\label{equ:rhodisc}
\end{align}
Both $\rho_{\mathrm{thin},0}$ and $\rho_{\mathrm{thick},0}$ share the total number density of the disc, $\rho_{\mathrm{disc},0}$, in a corresponding ratio of $1:0.12$.

In addition, we also model a two-axial power-law ellipsoid for halo stars as follows,
\begin{align}
\rho_{\mathrm{halo}}(R,z)=\rho_{\mathrm{halo}}(R_\odot,0)\left(\frac{R_\odot}{\sqrt{R^2+(z/q_{\mathrm{H}})^2}}\right)^{n_\mathrm{H}}
\end{align}
where the parameter $q_H$ represents the halo ellipticity. 
A value of $q_H<1$ indicates that the halo exhibits an oblate shape characterised by a flattened disc-like morphology.
Here we make an approximation, $z+z_\odot\approx z$, for the halo component since the solar height is negligible compared to the halo scale.
The local halo number density $\rho_{\mathrm{halo}}(R_\odot,0)$ (hereafter $\rho_{\mathrm{halo},0}$) is also normalised to the thin disc, 
which is appropriately scaled to match the distribution of stars in different components.
We consider the local density radio for three components as $\rho_{\mathrm{thin},0}:\rho_{\mathrm{thick},0}:\rho_{\mathrm{halo},0}=1:0.12:0.005$, allowing us to accurately simulate the distribution of stars across the different Galactic structures. Note that, for the following results, we will use only the relative ratios of these three components to evaluate the selection function of disc and halo WDs. 
In line with the approximation made for the disc profile, we also disregard the variation in the Galactic radial direction by setting $R\simeq R_\odot$ in the above equation. 
Consequently, we use a simplified density profile to simulate the stellar halo: 
\begin{align}
\rho_{\mathrm{halo}}(z)\simeq \rho_{\mathrm{halo},0}\left[
1+\left(\frac{z}{R_\odot q_{\mathrm{H}}}\right)^2\right]^{-\frac{n_\mathrm{H}}{2}}
\end{align}
Here we adopt $n_H=2.8$ and $q_H=0.64$ \citep{2008ApJ...673..864J}.
The $R_\odot$ represents the distance from our Sun to the Galactic centre set as 8,122~pc in this work.

\begin{figure}
\begin{center}
 \includegraphics[width=\columnwidth]{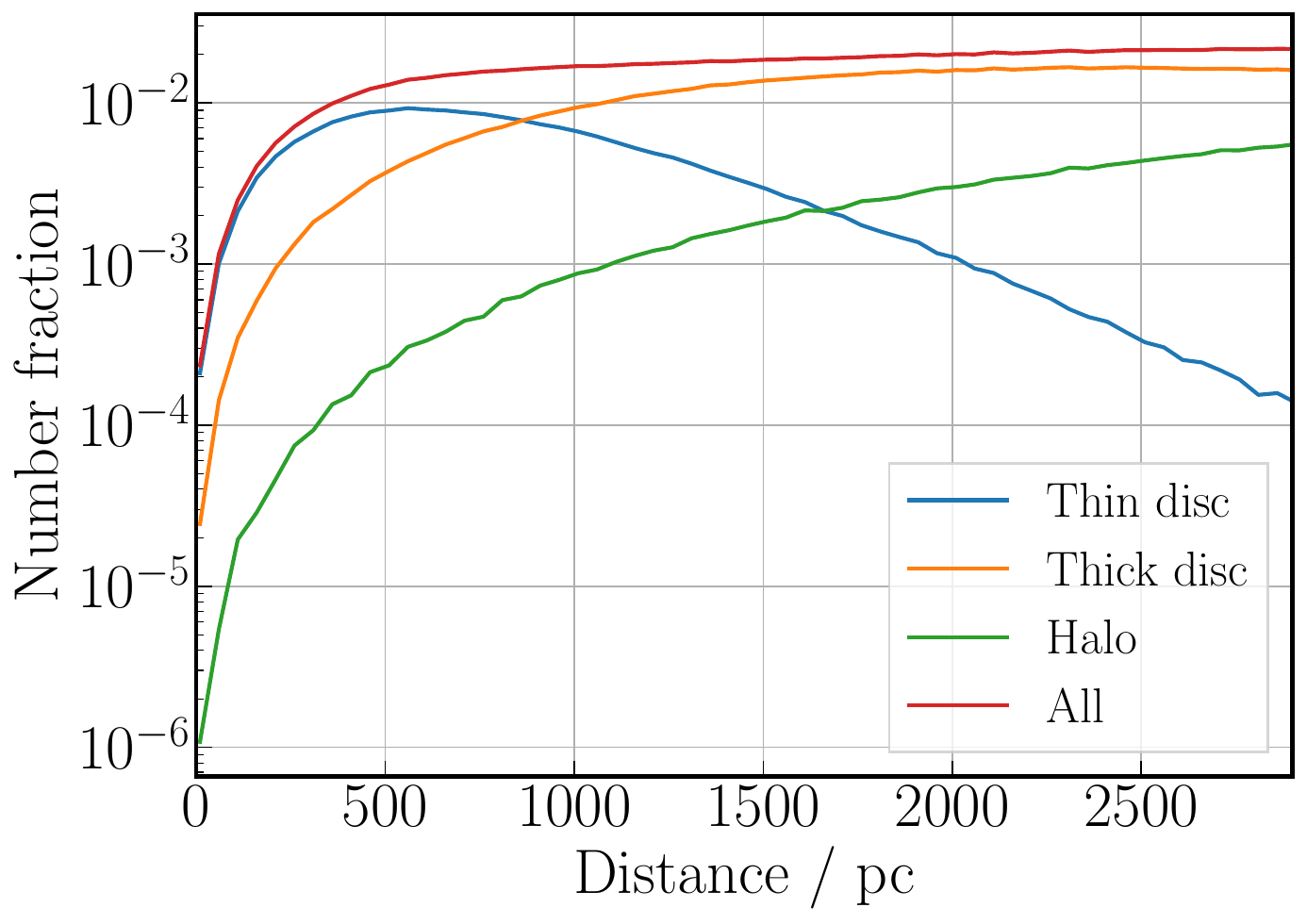}
 \caption{The 
 distance distribution of each of the different WD populations that belong to the Galactic thin and thick disc and the halo regions, respectively. 
 We compute these distributions using the mock catalogues generated assuming the standard Galactic model (see text for details). 
 Note that the distance is from the Sun. 
 The lines show the normalised numbers of each component in each distance bin, spaced by 50~pc. 
 The line for ``All'' (red line) is normalised to satisfy $\sum_i n(d_i)=1$, where $i$ denotes the $i$-th distance bin. 
 Other curves give a relative contribution to the total (All).
 All the lines are after the average of the number counts over the S82 region.
 }
 \label{fig:simu}
\end{center}
\end{figure}

In this paper, we generate $10^6$ stars (equivalently WDs)
up to 3~kpc distance in the region of S82.
We illustrate the distance distribution of our mock stars in Figure~\ref{fig:simu}.
The stars in the stellar halo show an almost no-decreasing distribution.
The stars in the thin disc dominate the sample within a few hundred parsecs, while the stars in the thick disc become the dominant role after that.
This comparison highlights the variations in number density among the different components and provides insight into their spatial distribution.
\subsubsection{Velocity distribution}
\label{ssec:Velocity}
\begin{table}
 \begin{center}
 \begin{tabular}{crrr}
  \hline
  \hline
  Velocity paras & Thin disc & Thick disc &  Stellar halo\\[2pt]
  $[{\rm km}~{\rm s}^{-1}]$\\[2pt] \hline
  $\langle U\rangle$
  & $-8.62$ & $-11.0$ & $-26.0$\\[2pt]
  $\langle V\rangle$ & $-20.04$ & $-42.0$ & $-199.0$\\[2pt]
  $\langle W\rangle$ & $-7.10$ & $-12.0$ & $-12.0$\\[2pt]
  $\sigma_U$ & 32.4 & 50.0 & 141.0\\[2pt]
  $\sigma_V$ & 23.0 & 56.0 & 106.0\\[2pt]
  $\sigma_W$ & 18.1 & 34.0 & 94.0\\[2pt]
  \hline
 \end{tabular}
 \end{center}
 \caption{The kinematic quantities adopted in our mock WD catalogue that includes the simulated  
 velocity distributions of the disc and 
 halo WDs. 
 Shown are the mean velocity and velocity dispersion for each component with respect to our Sun in the Milky Way Galaxy. 
 The parameters for the thin disc are obtained by \protect\cite{2009AJ....137.4149F}, while the ones for the thick disc and halo are from \protect\cite{2000AJ....119.2843C}.}
\label{tab:velosity}
\end{table}
\begin{figure}
\begin{center}
 \includegraphics[width=\columnwidth]{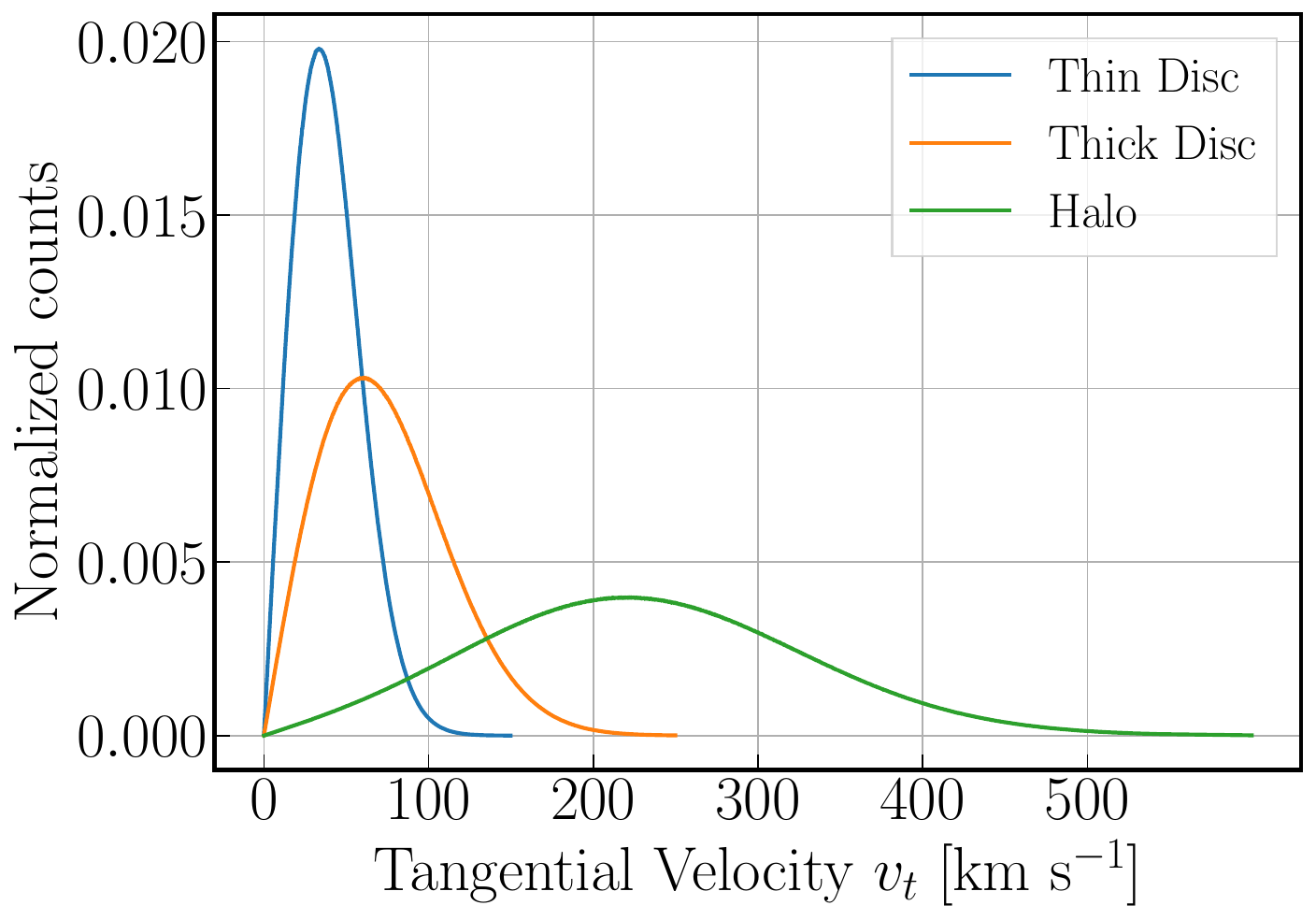}
 \caption{Similarly to the previous figure, this figure shows the tangential velocity distribution for the thin disc, thick disc and halo WDs in the mock catalogue. 
 The curve for each component has been normalised; hence, the relative height does not carry any information. 
 The distributions of the disc WDs and the halo WDs significantly differ at the large velocity end.
   }
 \label{fig:vtdistribution}
\end{center}
\end{figure}
In addition to the density profile, we also use the velocity distribution of stars (WDs).
For stars in the disc and halo regions, we assume a Gaussian distribution with mean and width given in Table~\ref{tab:velosity}, for the velocity components $U, V,$ and $W$ relative to the Sun. 
Here $U$, $V$ and $W$ are the velocity components in the Cartesian coordinates, defined in that $U$ is the component towards the Galactic centre, $V$ is the component along the Galactic rotation, and $W$ is the component perpendicular to the Galactic plane ($z$-direction).

We assign a simulated velocity to each star particle of the mock catalogue, constructed in the preceding section, assuming the velocity distribution for each component in Table~\ref{tab:velosity}.
Thus we simulate a representative distribution of
the positions and velocities of WDs in the disc and halo regions. 
Note that we use the \texttt{PYTHON} package \texttt{astropy} to perform the coordinate transformation of simulated velocity between 
the coordinate systems of (R.A., Dec.) and $(l,b)$. 

Figure~\ref{fig:vtdistribution} shows the simulated distribution of tangential velocities for disc and halo WDs.
The small, intermediate and large velocity WDs tend to be in the thin disc, thick disc and halo regions, respectively. 
In particular WDs with large velocities, $v_{\rm t}\gtrsim 200~{\rm km}~{\rm s}^{-1}$ are probably halo WDs since such WDs are rest with respect to the Galactic centre on average and should have large reflex motions with respect to the Sun that has a rotation velocity of $\simeq 230~{\rm km}~{\rm s}^{-1}$ with respect to the Galactic centre.
We can utilise these distinct velocity distributions to select the disc and halo WD candidates in our sample, 
which are also used to compute the discovery fraction as detailed in Section~\ref{ssec:DF}.

However, note that we cannot cleanly separate the thin and thick disc WDs based solely on tangential velocity. 
Although the study by \citetalias{2011MNRAS.417...93R} proposed a statistical distinction between thin and thick discs for WDs, we cannot identify the two WD populations on an individual basis.
This is due to the absence of additional information such as radial velocity and metallicity. 
Consequently, we will not attempt to disentangle the thin and thick discs or calculate their luminosity functions respectively. 
Instead, we will treat the thin and thick discs as a single entity with a unified density profile like 
Eq.~(\ref{equ:rhodisc}).
\subsubsection{Expected fraction of disc and halo WDs}
\label{ssec:fraction_haloWD}
\begin{figure}
\begin{center}
\includegraphics[width=0.9\columnwidth]{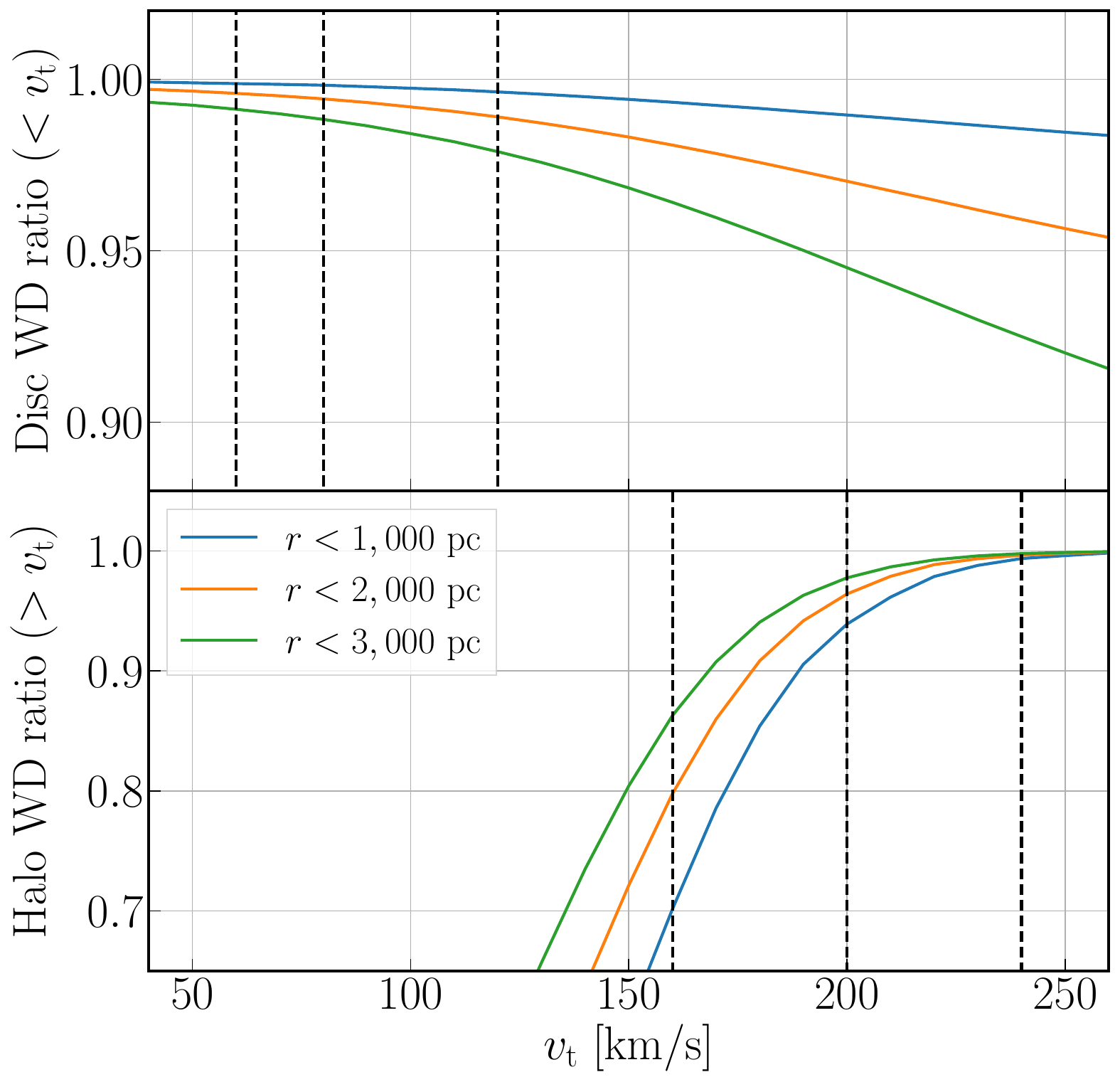}
 \caption{The fraction of WDs belonging to each component is shown for different tangential velocity cuts with distance limits of 1,000, 2,000, and 3,000~pc, respectively
 The upper panel shows the fraction of stars with an upper-velocity cut,
 while the lower panel shows one of the stars with a lower-velocity cut.
 The black dashed lines indicate the velocity cuts of 60, 80 and 120~km~s$^{-1}$, respectively, in the upper panel, while for 160, 200, and 240~km~s$^{-1}$ in the lower panel, which are the selection criteria for our disc and halo WD samples in Section~\ref{sec:WDLF}. 
 These lines infer the contamination from each other component, based on the tangential velocity cuts.}

 \label{fig:frac}
\end{center}
\end{figure}
As we show in Figures~\ref{fig:simu} and \ref{fig:vtdistribution}, we have the mock catalogues for the number density and velocity distributions of WDs in the disc and halo regions. Using the mock, we can compute the expected fraction of dick or halo WDs among all WDs as a function of distance and tangential velocity. 
Figure~\ref{fig:frac} shows the results. 
The figure clearly shows that the tangential velocity gives a useful way to disentangle disc and halo WDs, while the distance is not so useful in comparison.
To be more precise, if we select WDs by the velocity cut of $v_{\rm t}<120$~km~s$^{-1}$, most of the WDs (more than $97\%$) are very likely to be disc WDs. If we select large-velocity WDs with $v_{\rm t}>160$~km~s$^{-1}$, more than 70\% of WDs are halo ones. We can increase The fraction of halo WDs by focusing on more distant WDs. 
On the other hand, WDs with intermediate velocities such as $120\lesssim v_{\rm t}/[{\rm km}~{\rm s}^{-1}]\lesssim 160$ have non-negligible contamination to 
either of dick or halo WDs from the other population. Given these results, we will use cuts of $v_{\rm t}$ to construct samples of disc and halo WDs from our sample in Section~\ref{sec:WDLF}. 

\subsubsection{Representativeness of WDs in the Stripe~82}

Our research focuses on a specific region of the sky, covering approximately 165 square degrees
for the Stripe~82. Given that the entire sky spans over 40,000 square degrees, a critical consideration is whether the WDs identified in this relatively small area are representative of the broader WD population.

\begin{figure}
\begin{center}
\includegraphics[width=\columnwidth]{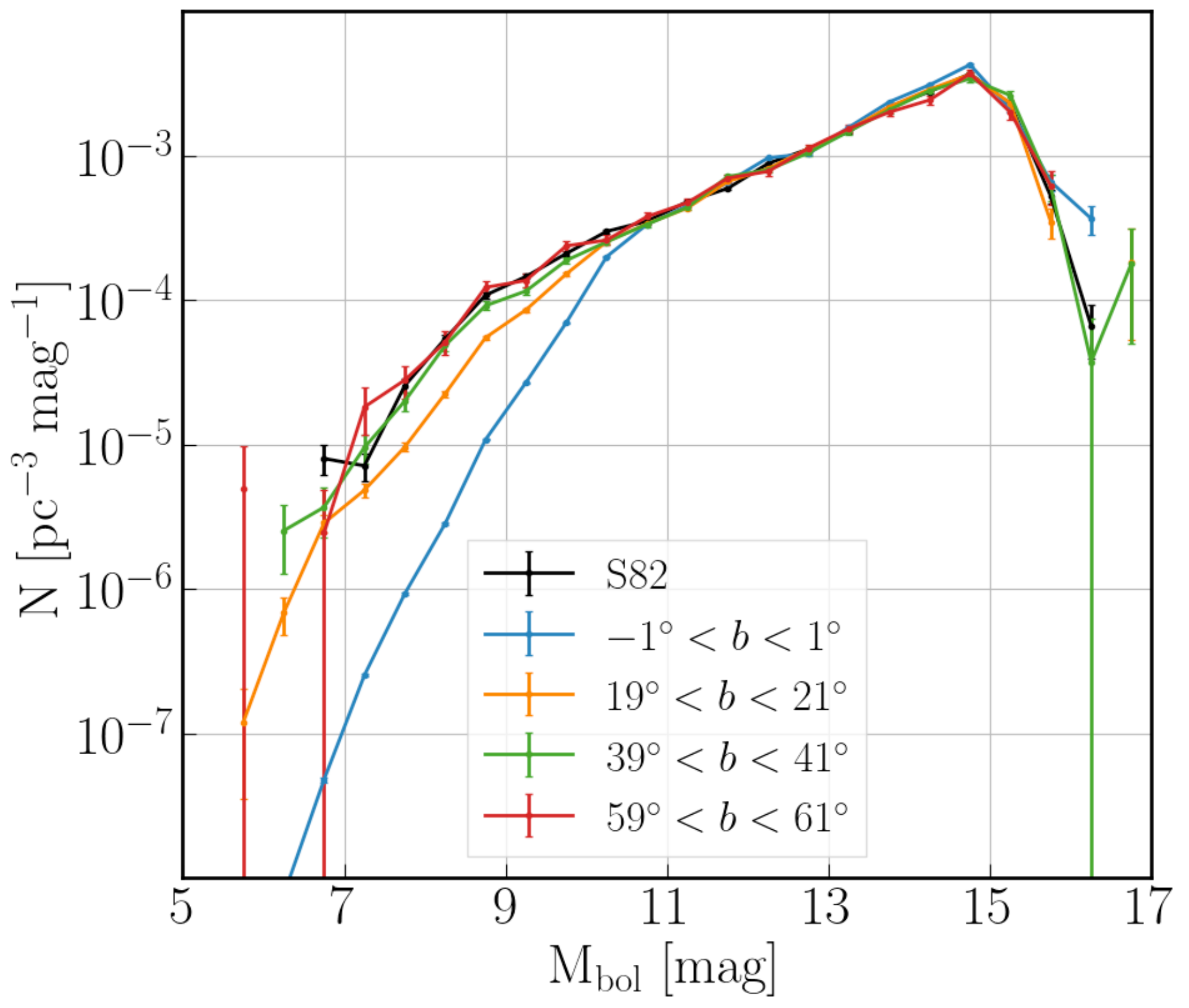}
 \caption{Luminosity functions of the mock WDs from different regions in the Galaxy.
 The subsample slice with Galactic latitude (b) varies from 0° to 60° with a width of 2°, showing that the results remain consistent across these latitudes. 
 Note that our mock data is limited to a distance of 5,000 pc, which does not provide complete coverage for absolute magnitudes brighter than 10.5 mag. 
 Therefore, the consistency in results is only for magnitudes fainter than 10.5 mag. }
 \label{fig:simu_comp}
\end{center}
\end{figure}

To address this concern, we use mock catalogues of WDs using the standard model of the Milky Way (also see below). We simulated the spatial and velocity distributions for each WD using the standard model 
for the thin and thick discs and the halo region of the Milky Way. We also adopt the luminosity function 
of WDs obtained from {\it Gaia} DR3 \citetalias{2021A&A...649A...6G}, and assign 
an absolute magnitude to each WD in a way that the population is consistent with the luminosity function 
in the disc and halo regions. Then we can compute the apparent magnitude for each WD from the simulated 
distance. 
Using the above mock catalogue, we simulated the same selection of WD as used in the actual data. 
To assess the consistency of the luminosity functions of mock WDs across different regions of the Galaxy, we analyse 4 subsamples with Galactic latitude (b) varying from $0^{\circ}$ to $60^{\circ}$ in slices of $2^{\circ}$ width, while the Galactic longitude (l) does not affect the results.
Figure \ref{fig:simu_comp} illustrates that the results remain consistent across these latitudes. 
It is important to note that our mock data is limited to a distance of 5,000 pc , which does not provide complete coverage for absolute magnitudes brighter than 10.5 mag, since including data beyond this distance in the Galactic disc would result in an extremely large dataset.
Consequently, the observed consistency pertains to magnitudes fainter than 10.5 mag.

Therefore, we have confirmed that the WDLF for the S82 region is representative of the underlying true WDLF in the MW. 

\subsection{Discovery fraction}
\label{ssec:DF}
As shown in Figure~\ref{fig:frac}, the tangential velocity of each WD candidate, estimated from the photometric distance ($r$) and the proper motion ($\mu$), gives a useful quantity to separate disc and halo WDs. Please do not confuse the notation of distance, $r$, with the $r$-band apparent magnitude in this section; we use ``$r$'' to denote the distance throughout this and the following subsections. 
However, in practice, we need to employ a finite range of cuts in $v_{\rm t}$ to minimise the contamination.
For small velocity $v_{\rm t}\lesssim 40~{\rm km}~{\rm s}^{-1}$, faint main-sequence stars could contaminate the WD sample, so we employed the cut of $v_{\rm t}>40~{\rm km}~{\rm s}^{-1}$ to define the parent WD sample (see Figure~\ref{fig:RPMmodel},
Table~\ref{tab:WDnum} and Section~\ref{ssec:summary_wd_sample}). For WDs with $120\lesssim v_{\rm t}/[{\rm km}~{\rm s}^{-1}]\lesssim 160$, it is difficult to cleanly separate disc and halo WDs.

Hence, to correct for the selection effect of disc or halo WDs due to the cuts of $v_{\rm t}$, we use 
the ``discovery fraction'', which was first introduced in \citetalias{2006AJ....131..571H}.
According to the results in Figure~\ref{fig:frac}, we will employ an adequate range of cuts in $v_{\rm t}$ to select disc or halo WDs in Section~\ref{sec:WDLF}.
The discovery fraction for ``disc'' or ``halo'' WDs is defined as

\begin{align}
\chi_{\rm{disc}}(r)=\int^{v_{\rm u}(r)}_{v_{\rm l}(r)}\!\!\mathrm{d}v_{\rm t}~
P_{\rm disc}(v_{\rm t})~\nonumber\\
\chi_{\rm{halo}}(r)=\int^{v_{\rm u}(r)}_{v_{\rm l}(r)}\!\!\mathrm{d}v_{\rm t}~
P_{\rm halo}(v_{\rm t})
\end{align}
Where $P_{\rm disc}(v_{\rm t})$ or $P_{\rm halo}(v_{\rm t})$ is the probability function of $v_{\rm t}$
for disc or halo WDs as given in Figure~\ref{fig:vtdistribution}, respectively. 
The lower and uppercuts in the integration are
\begin{align}
v_{\rm u}(r)&={\rm min}(v_{\rm uppercut},4.74\times r\times \mu_{\rm min})~\nonumber\\
v_{\rm l}(r)&={\rm max}(v_{\rm lowercut},4.74\times r\times \mu_{\rm max})~
\end{align}
where $v_{\rm uppercut}$ and $v_{\rm lowercut}$ are the upper and lower cuts of $v_{\rm t}$ used to select 
disc or halo WDs from the sample which we have illustrated in Figure~\ref{fig:frac} by dashed lines.
$\mu_{\rm max}$ is the maximum cut of the proper motion measurement: $\mu_{\rm max}=130~{\rm mas}~{\rm yr}^{-1}$ (see Figure~\ref{fig:comp_pm}), where the numeric factor $4.74$ is a conversion factor (see Eq.~\ref{eq:vt_def}). 
$\mu_{\rm min}$ is the minimum cut used in the proper motion measurement for which we employ $4\sigma$ detection: 
$\mu_{\rm min}=4\sigma_\mu(i)$, where $\sigma_\mu(i)$ is the 1-$\sigma$ statistical error of the proper motion measurement for WDs at ``$i$'' apparent magnitude as given in Eq.~(\ref{eq3}) (see Section~\ref{sssec:lowerlimit} for the details). 
Although $\int_{0}^\infty\mathrm{d}v_{\rm t}~P(v_{\rm t})=1$, the finite-range integration gives $0\le \chi(r)\le 1$. Hence we will multiply the estimator of the luminosity function by $1/\chi_{\rm disc/halo}$ to correct for the selection effect or incompleteness due to the cut of $v_{\rm t}$. 

\subsection{Maximum volume density estimator}
\label{ssec:MaxV}
We use the maximum volume density estimator \citep{1968ApJ...151..393S} \citep[also see][]{1976ApJ...207..700F} to estimate the luminosity function of WDs.
We combine the method in \citetalias{2006AJ....131..571H} and \citet{2015MNRAS.450.4098L} with an estimator to correct for the incompleteness due to the intrinsic faint WDs, 
using the discovery fraction defined in Section~\ref{ssec:DF}.

We use the modified maximum volume estimator for the LFs for disc or halo WDs following \citetalias{2011MNRAS.417...93R}.
For each WD candidate, say the $k$-th WD here, we compute the ``maximum volume'' in that the object can be detected by our magnitude cuts:
\begin{align}  
    V_{{\rm max},k}=\Omega_A\int_{r_{{\rm min},k}}^{r_{{\rm max},k}}r^2\mathrm{d}r~\frac{\rho_{\rm disc}(r)}{\rho_{\odot,{\rm disc}}}\chi_{\rm disc}(r)~\nonumber\\
    V_{{\rm max},k}=\Omega_A\int_{r_{{\rm min},k}}^{r_{{\rm max},k}}r^2\mathrm{d}r~\frac{\rho_{\rm halo}(r)}{\rho_{\odot,{\rm halo}}}\chi_{\rm halo}(r)
\label{eq:Vmax}
\end{align}
where $\Omega_A=0.05026$ is the solid angle of the S82 region, and $\chi_{\rm disc/halo}(r)$ is the discovery fraction at distance $r$ for the given cuts of tangential velocity that are used to select disc and halo WDs. 
The function $\rho_{\rm disc/halo}(r)/\rho_\odot$ takes into account the radial profile probability of the $k$-th WD for being in the disc and halo region
(see Section~\ref{ssec:density}).
We use the radial function normalised at the solar position following \citetalias{2011MNRAS.417...93R} so that we can compare our results with the previous works, and also take into account the dependence of the Galactic latitude ($b$) of each WD candidate. 
The integral lower and upper limits of distance for the $k$-th WD, $r_{{\rm min},k}$ and $r_{{\rm max},k}$, are determined by the detection limits of each filter as
\begin{align}
    r_{{\rm min},k}=r_{0,k}\times \mathrm{max}[10^{0.2(m_{\mathrm{min,}a}-m_{(k)a,{\rm obs}})}]\nonumber\\
    r_{{\rm max},k}=r_{0,k}\times \mathrm{min}[10^{0.2(m_{\mathrm{max,}a}-m_{(k)a,{\rm obs}})}]
\end{align}
where $a$ denotes the $a$-th HSC filter, i.e. $a=\{g,r,i,z,y\}$, $m_{{\rm min},a}$ is the lower cut of apparent magnitude for the $a$-th filter, which is $(18.27, 18.68, 19.00, 19.08, 19.07)$ used
in this work, respectively.
$m_{{\rm max},a}$ is the upper cut, and $m_{(k)a,{\rm obs}}$ is the observed magnitude of the $k$-th WD candidate in the $a$-th filter, which is $(26.58, 25.18, 24.00, 24.16, 24.43)$, respectively.
The maximum or minimum function chooses the maximum or minimum number from those of the $grizy$ filters. 
$r_{0,k}$ is the photometric distance for the $k$-th WD candidate, as discussed in Section~\ref{ssec:WDmodel}.

Then we can estimate the luminosity function of WDs in the absolute magnitude range $[M,M+\mathrm{d}M]$ as
\begin{align}
    \Phi(M)\mathrm{d}M=\sum_{k; M\le M_k\le M+\mathrm{d}M}\frac{1}{c(i_{(k)})}\frac{w_k}{V_{{\rm max},k}},
\end{align}
where the summation runs over all WD candidates that reside in the magnitude range $[M,M+\mathrm{d}M]$ after the velocity cuts, 
and $c(i_{(k)})$ is the matching completeness between the HSC and S82 catalogues for the proper motion measurements, which is given as a function of the apparent magnitude in the HSC $i$ band shown in Figure~\ref{fig:comp}. 
The weight $w_k$ takes into account the fractional probability of the $k$-th object for being in the absolute magnitude bin
from each of the fitting solutions of the blackbody or the H or He atmosphere WD model in Section~\ref{ssec:weights}.

Assuming the Poisson noise in the number counts, we can assign the statistical errors to the luminosity function in each magnitude bin as
\begin{align}
    \sigma^2(M)=\sum_{k; M\le M_k\le M+\mathrm{d}M}\frac{1}{c^2(i_{(k)})}\frac{w^2_k}{V^2_{{\rm max},k}}.
\end{align}
%

%%%%%%%%%%%%%%%%%%%%%%%%%%%%%%%%%%%%%%%%%%%%%%%%%%
\section{White dwarf luminosity function}
\label{sec:WDLF}
In this section, we show the main results of this paper, which are estimating the luminosity functions for WDs in the disc and halo regions from 5,080 WD candidates in our catalogue.

\subsection{LFs for disc WDs}
\label{ssec:lowV}
\begin{figure}
\begin{center}
\includegraphics[width=\columnwidth]{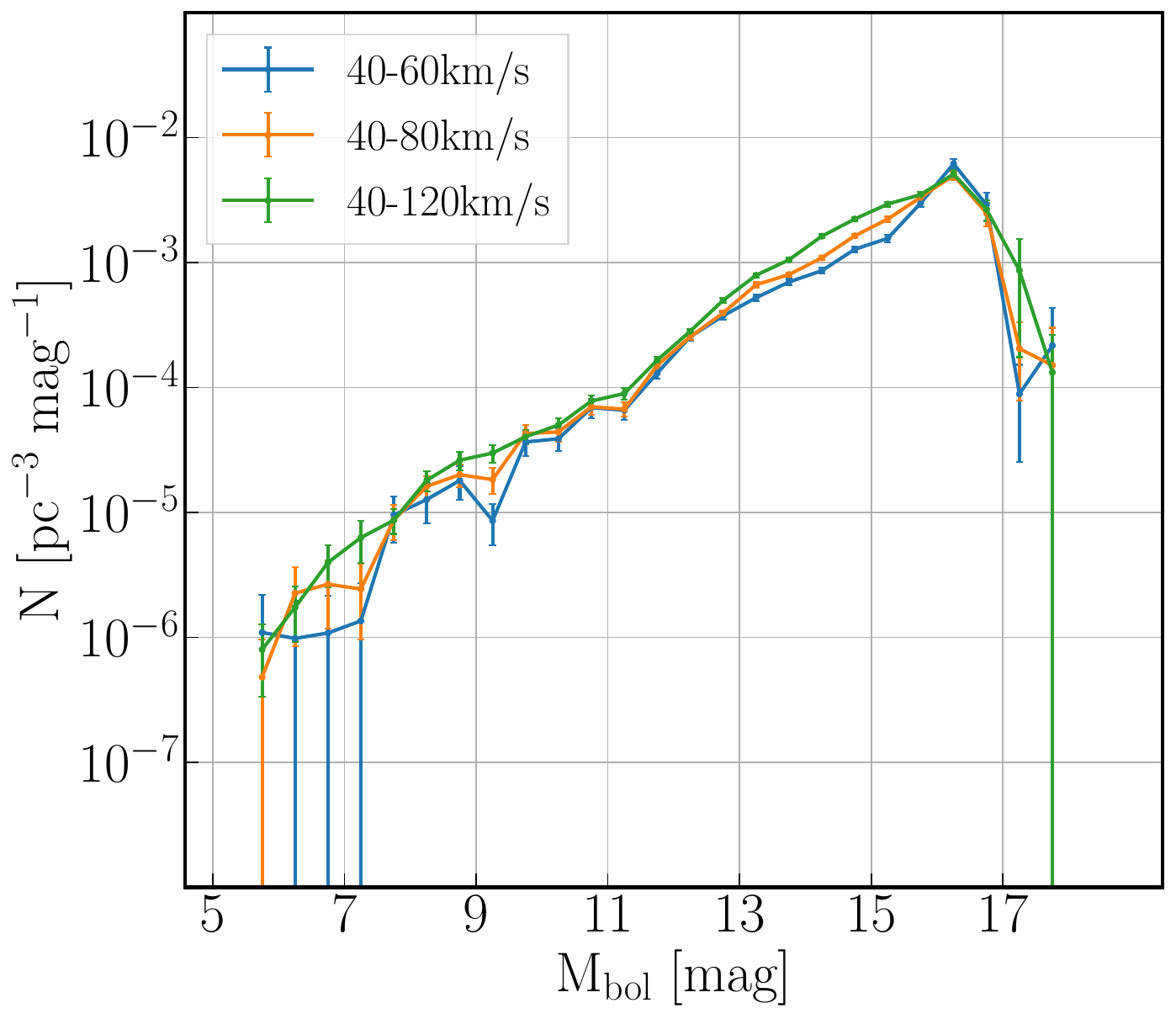}
 \caption{The luminosity function of the disc WD sample from our catalogue.
 We show a 0.5-magnitude-bin density profile for three different cuts of the tangential velocity ($v_{\rm t}$) as denoted by the legend.
 }
 \label{fig:disc_LF}
\end{center}
\end{figure}
As illustrated in Figure~\ref{fig:frac}, the WD candidates with tangential velocity $v_{\rm t}\lesssim 120~{\rm km}~{\rm s}^{-1}$ are predominantly associated with disc WDs.
In Figure~\ref{fig:disc_LF}, we present the  luminosity function of disc WDs using three different cuts of $v_{\rm t}$:
$40<v_\mathrm{t}<60$~km~s$^{-1}$, $40<v_\mathrm{t}<80$~km~s$^{-1}$, and $40<v_\mathrm{t}<120$~km~s$^{-1}$, respectively.
The corresponding catalogues contain 1,930, 3,022 and 4,217 objects, respectively. 
The integrated number density is $(9.15 \pm 1.08) \times 10^{-3}$~pc$^{-3}$, $(9.33 \pm 0.89) \times 10^{-3}$~pc$^{-3}$, and $(1.109 \pm 0.117) \times 10^{-2}$~pc$^{-3}$. 
These densities are mainly contributed by faint objects with bolometric magnitudes between 15 and 16.5.
The figure shows that the choice of velocity cuts does not significantly affect the results of the luminosity function. 
This outcome is expected because the fraction of disc WDs in our sample is not heavily contaminated by halo WDs, even if we assume a large distance of 3,000 pc, which most objects in our sample cannot reach, shown in the upper panel of Figure~\ref{fig:frac}. 

\begin{figure}
\begin{center}
\includegraphics[width=\columnwidth]{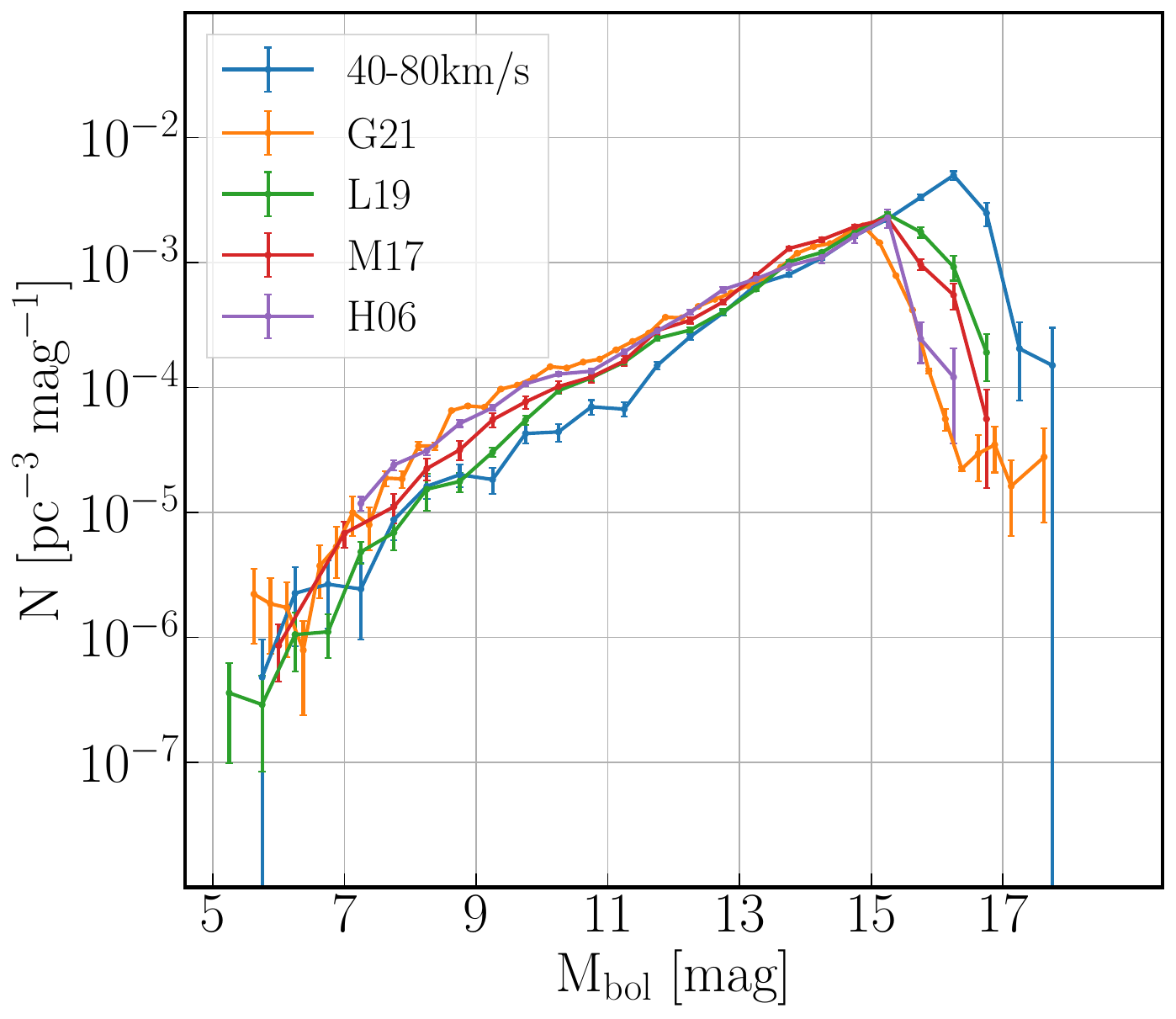}
 \caption{We compare our LF result for the disc WD sample with those from previous studies in 
 \citetalias{2021A&A...649A...6G}}, \citetalias{2019MNRAS.482..715L}, \citetalias{2017AJ....153...10M} and \citetalias{2006AJ....131..571H}.
 We choose the tangential velocity cut of 40--80~km~s$^{-1}$ here.
The results from \citetalias{2021A&A...649A...6G} did not assign tangential velocity limits to the WD samples. 
Instead, they utilised a complete sample of WDs from {\it Gaia} EDR3 within 100 pc to derive the luminosity function.
Our sample displays a higher peak at the fainter magnitude $M_{\mathrm{bol}}=16.25$.

 \label{fig:disc_Comparison1}
\end{center}
\end{figure}

\begin{figure}
\begin{center}
\includegraphics[width=\columnwidth]{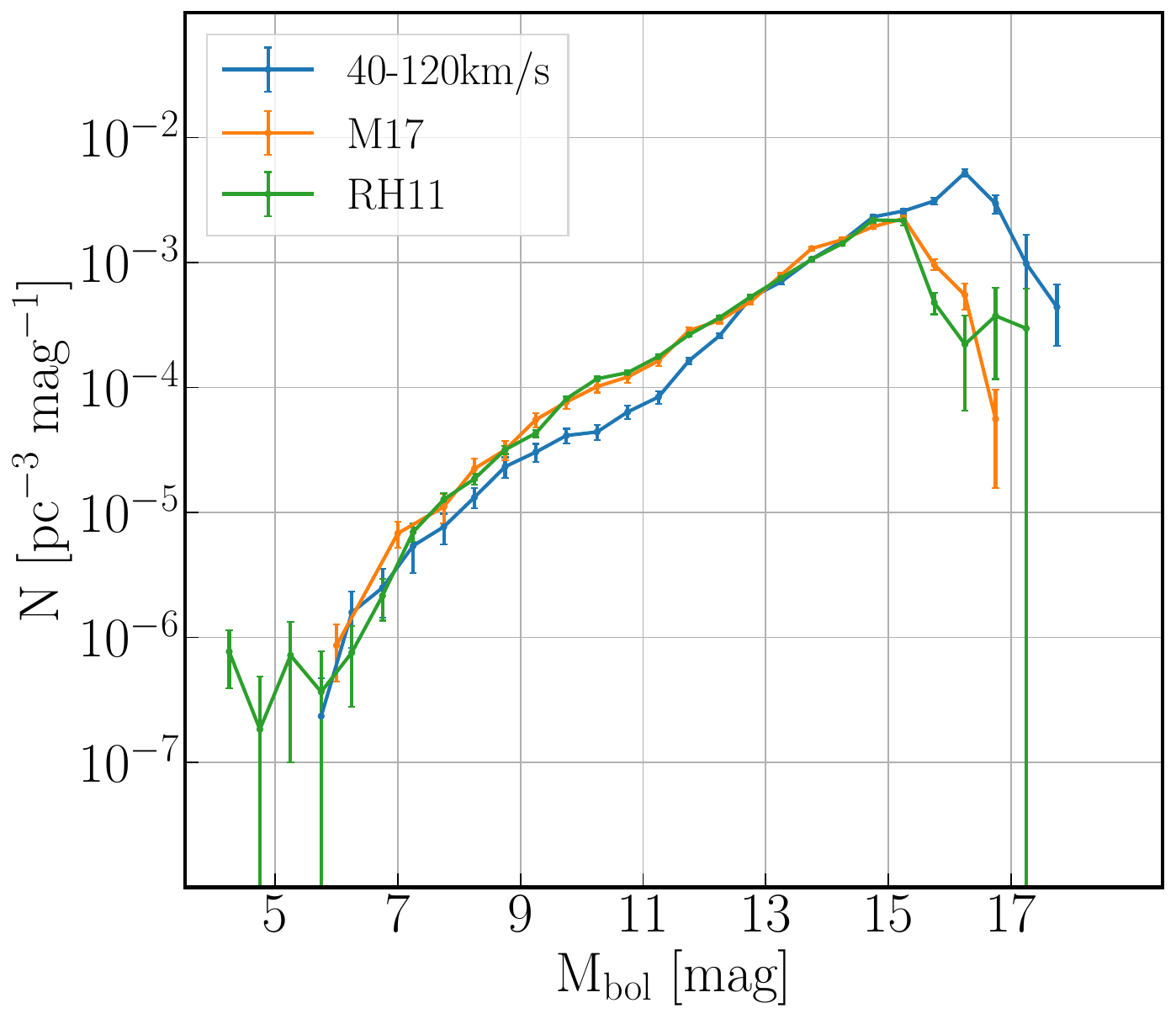}
 \caption{Similarly to the previous figure, we here compare our LF for the disc WDs with those from previous studies by \citetalias{2017AJ....153...10M} and \citetalias{2011MNRAS.417...93R}.
 In this comparison, we extended the tangential velocity limits to 120~km~s$^{-1}$.
 The overall features of the WDLF remained unchanged.
 }
 \label{fig:disc_Comparison2}
\end{center}
\end{figure}

We compare our WDLF results with those from previous studies, including \citetalias{2006AJ....131..571H}, \citetalias{2011MNRAS.417...93R}, \citetalias{2017AJ....153...10M}, \citetalias{2019MNRAS.482..715L} and 
\citetalias{2021A&A...649A...6G}.
All five previous works were based on large sky-area photometric surveys, while our study focuses on a smaller region of only 165~deg$^2$.
We present comparisons for a 40--80~km~s$^{-1}$ sample in Figure~\ref{fig:disc_Comparison1} and 40--120~km~s$^{-1}$ in Figure~\ref{fig:disc_Comparison2}. 
Please note that the results from \citetalias{2021A&A...649A...6G} did not assign tangential velocity limits to the WD samples. 
Instead, they utilised a complete sample of WDs from {\it Gaia} EDR3 within 100 pc to derive the luminosity function.
Overall, our results are in good agreement with the previous works for the bright part ($\mathrm{M_{bol}}<15$), with a minor insufficiency observed around magnitude 11. 
This deficit of WD candidates can be attributed to a few factors.
One contributing factor is the lack of enough bright WD candidates.
The specific design of our survey region, which is directed away from the Galactic plane, aims to minimise the effects of dust and bright stars from the Galactic plane and centre. 
However, this also leads to a reduction in the number of stars belonging to the Galactic disc, including nearby (therefore relatively bright) WDs.
Another factor is the limitation of the photometric fitting process. 
The lack of UV band data in our observations can introduce uncertainty, particularly for high-temperature WDs with temperatures above 10,000~K with potential strong absorption lines, which contributes to the number density for bolometric magnitude brighter than 13. 
UV bands are more sensitive to the spectra of such WDs, which cover the turnover in the blackbody spectra.

Despite that, the deep and faint observations have allowed us to identify a significant peak at ${M}_{\mathrm{bol}}=16.25$ in the WDLF, with a density of $(4.98 \pm 0.39) \times 10^{-3}$~pc$^{-3}$~mag$^{-1}$.
Furthermore, there is still a notable density of WDs with ${M}_{\mathrm{bol}}=17$. 
If this trend truly holds at such faint magnitudes, it suggests the discovery of the oldest WDs in our Milky Way Galaxy, which may have aged as old as the Galaxy itself.
The estimated cooling times for WDs, based on theoretical models, present a puzzling result, as they seem to exceed the age of the Galactic disc, which is around 9 billion years. 
According to the current models, DA WDs with a surface gravity of log$g=8.0$ would take 10.4 billion years to reach ${M}_{\mathrm{bol}}=16.5$, while for DB WDs, the cooling time would be 9.5 billion years. 
The large number of faint WD candidates in the disc is unexpected given their long cooling times.
One possible explanation for this discrepancy is that the faint WD candidates may exhibit different properties, such as mass and surface gravity, compared to most WDs. 
However, this explanation does not account for the total number of WDs fainter than 16 mag, which is 289 within the tangential velocity range of 40 to 80~km~s$^{-1}$. 
It is unlikely that all of these candidates can be attributed solely to their specific properties.
Another intriguing possibility is that the WDs observed in the faint end are ancient remnants that provide valuable insights into the early stages of Galactic merger events. 
These ancient remnants may be associated with structures like the dwarf galaxy ``Gaia sausage'' \citep{2018Natur.563...85H}. 
Figure 2 in \cite{2019NatAs...3..932G} also highlights a significant population of ancient objects within the thick disc, persisting for over 10 Gyr.
Such a scenario could explain the presence of a large number of faint WDs in the Galactic disc, as they would be relics of past merger events. 
We will discuss these faint WD candidates in detail in Section~\ref{sec:faintWD}.

\subsection{LFs for halo WDs}
\label{ssec:highV}
On the other hand, we can preferentially select halo WDs by imposing the velocity cut of $v_\mathrm{t}>200$ km s$^{-1}$.
As displayed in the lower panel of Figure~\ref{fig:frac}, the 240~km~s$^{-1}$ threshold can almost avoid all the contamination from the disc but will lead to a really low number of objects passing it.
The 160~km~s$^{-1}$ can significantly contaminate the halo WD pool by around 30\%, while the 200~km~s$^{-1}$ is acceptable for contamination and still keep enough objects.
The upper limit of 500~km~s$^{-1}$ comes from the Galactic escape velocity of $528^{+24}_{-25}$~km~s$^{-1}$ \citep{2019MNRAS.485.3514D} to remove the unbound objects in the Galaxy.
\begin{figure}
\begin{center}
\includegraphics[width=\columnwidth]{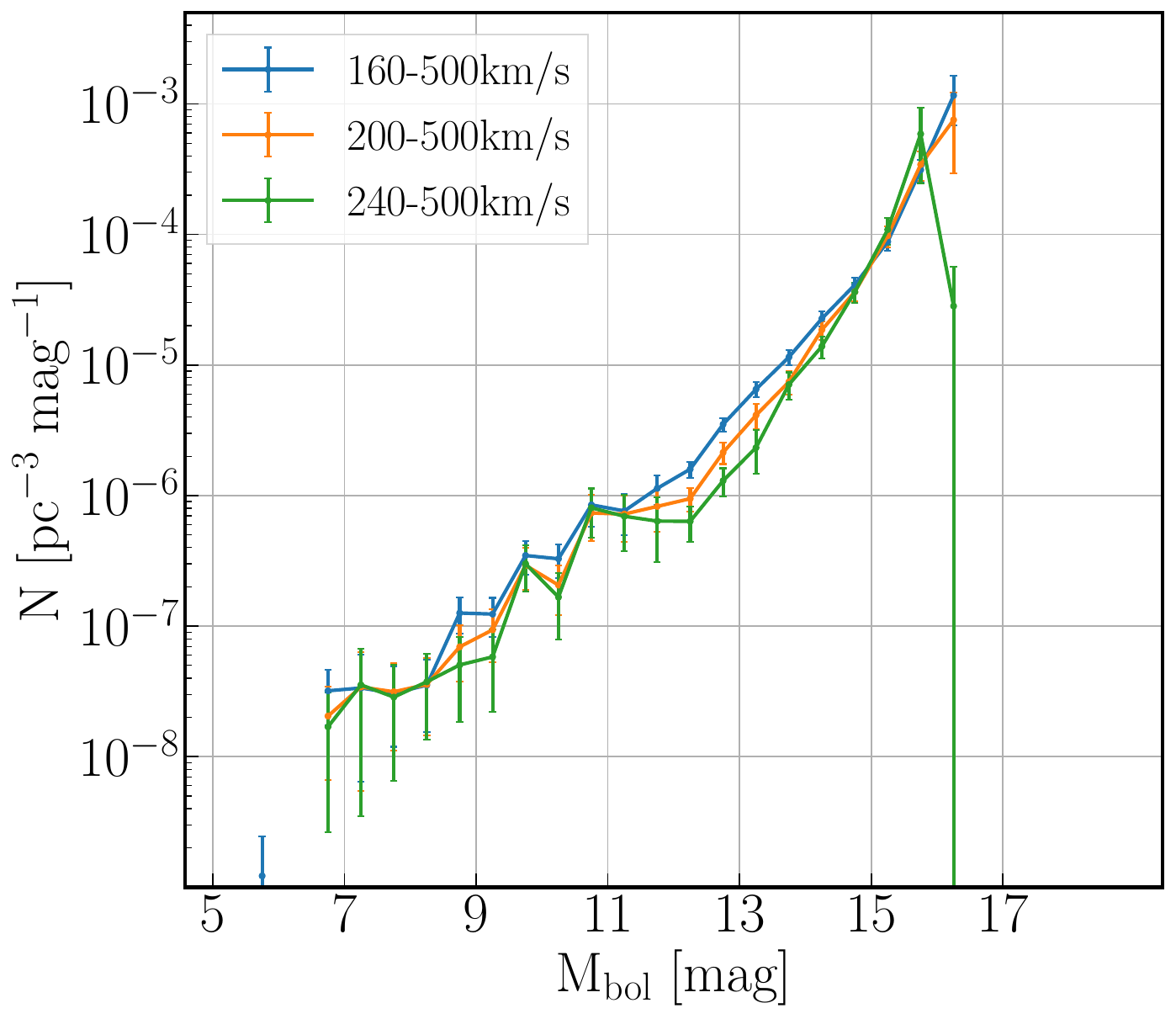}
 \caption{The luminosity function of the halo WD sample from our catalogue.
 We show a 0.5-magnitude-bin density profile for three different cuts of the tangential velocity as shown by the legend. 
 }
 \label{fig:halo_LF}
\end{center}
\end{figure}

Figure~\ref{fig:halo_LF} shows the luminosity function of halo WDs for three different cuts of $v_{\rm t}$: 
$160<v_\mathrm{t}<500$~km~s$^{-1}$, $200<v_\mathrm{t}<500$~km~s$^{-1}$, and $240<v_\mathrm{t}<500$~km~s$^{-1}$. 
The corresponding catalogues used for these analyses contain 432, 235, and 146 objects, respectively.
The selection of velocity cuts has a significant impact on the results. 
The 160~km~s$^{-1}$ sample includes a large number of objects from the velocity tail of the thick disc, which can lead to contamination in the halo luminosity function.
On the other hand, the stringent cut 240~km~s$^{-1}$ results in a smaller number of objects contributing to the luminosity function.
The total densities derived from these samples are: $(8.26 \pm 2.79) \times 10^{-4}$~pc$^{-3}$, 
$(6.34 \pm 2.89) \times 10^{-4}$~pc$^{-3}$, 
and $(3.95 \pm 2.06) \times 10^{-4}$~pc$^{-3}$, respectively. 
\begin{figure}
\begin{center}
\includegraphics[width=\columnwidth]{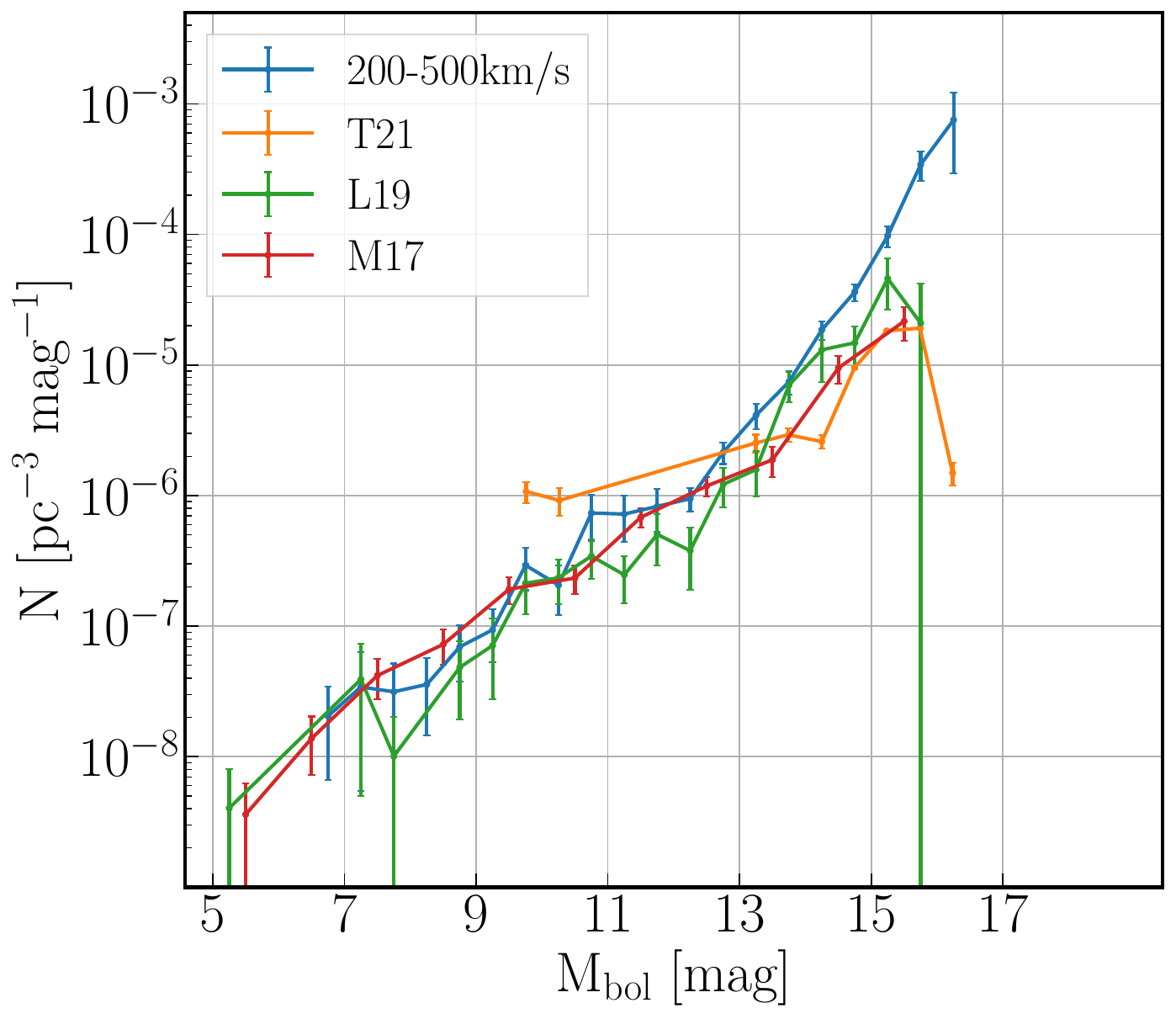}
 \caption{Comparison of our LF of halo WDs with those from previous studies by \citetalias{2021MNRAS.502.1753T}, \citetalias{2019MNRAS.482..715L} and \citetalias{2017AJ....153...10M}.
 The results from \citetalias{2021MNRAS.502.1753T} also did not assign tangential velocity. 
 They utilised a halo WD sample within 100 pc from {\it Gaia} DR2.
 Our result shows a similar profile to the works by \citetalias{2019MNRAS.482..715L} and \citetalias{2017AJ....153...10M}, while there is a much higher peak at the faint end thanks to our deep observation.}

 \label{fig:halo_Comparison1}
\end{center}
\end{figure}
\begin{figure}
\begin{center}
\includegraphics[width=\columnwidth]{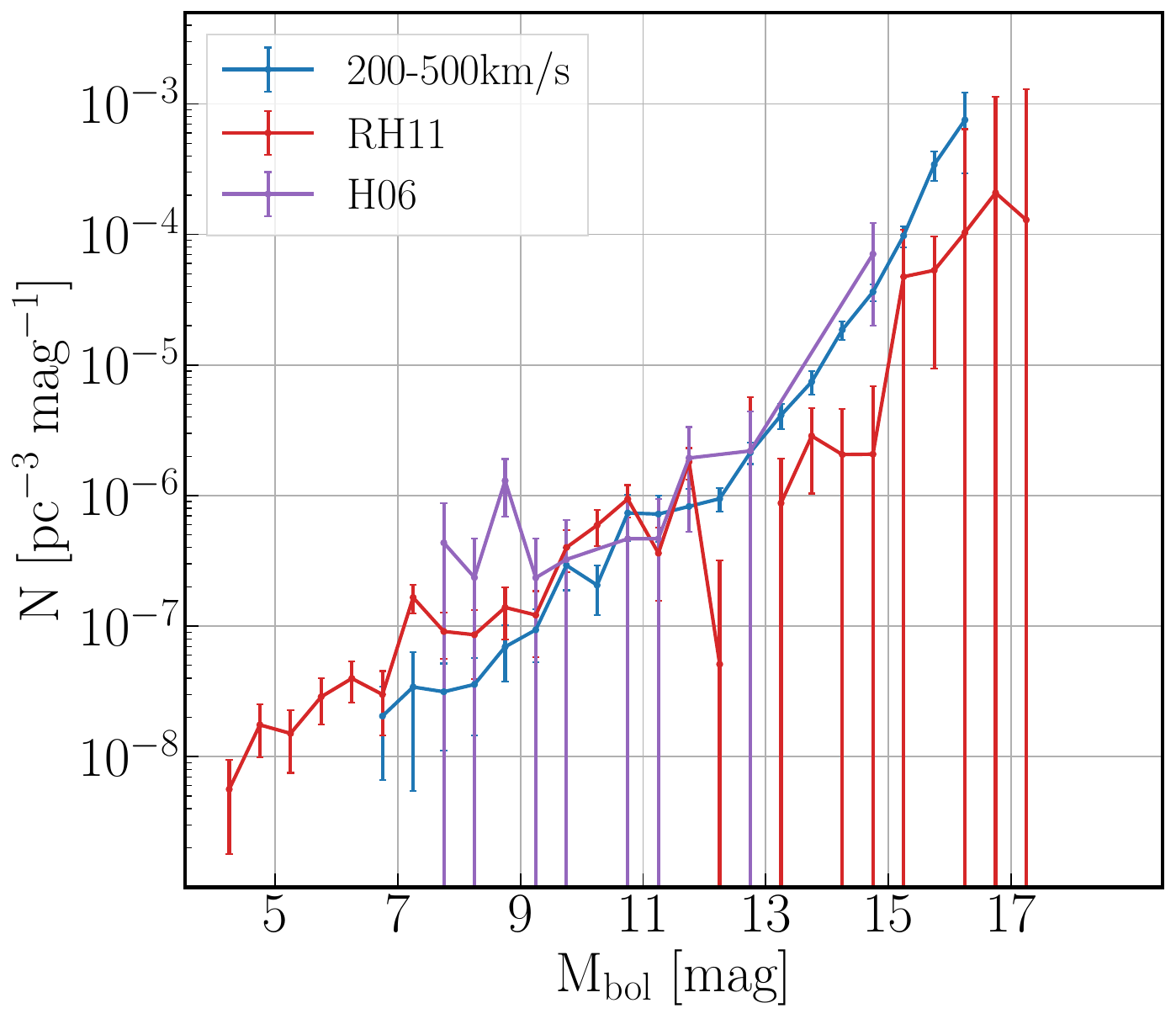}
 \caption{Similarly to the previous figure, we compare our LF of halo WDs with those from previous studies by \citetalias{2011MNRAS.417...93R} and \citetalias{2006AJ....131..571H}.
 Since the other two results have the larger errorbars, we show this comparison separately from the previous figure, for the sake of clarity. 
 }
 \label{fig:halo_Comparison2}
\end{center}
\end{figure}

We compare our results with previous works in Figures~\ref{fig:halo_Comparison1} and \ref{fig:halo_Comparison2}, focusing on the high-velocity samples within the tangential velocity range of 200~km~s$^{-1}$ to 500~km~s$^{-1}$.
The work by \citetalias{2006AJ....131..571H} is limited to only 18 objects with $v_\mathrm{t}>200$ km s$^{-1}$, while \citetalias{2011MNRAS.417...93R}, \citetalias{2017AJ....153...10M} and \citetalias{2021MNRAS.502.1753T} have 93, 135 and 95 objects, respectively, which is about half the size of our sample.
The results from \citetalias{2021MNRAS.502.1753T} also did not assign tangential velocity. 
They employed a halo WD sample within 100 pc from {\it Gaia} DR2.
The number density was obtained by the classic estimator $N/V$ rather than the maximum volume density estimator since they considered their sample as complete.
Our result for the high-velocity sample agrees well with previous works, particularly with the work by \citetalias{2019MNRAS.482..715L}.
However, a notable difference is observed at the faint end of the luminosity function.
Thanks to the inclusion of the faintest bin, the integrated density of our sample is much higher than $(5.291 \pm 2.717) \times 10^{-5}$ pc$^{-3}$ in \citetalias{2019MNRAS.482..715L} and $(3.5 \pm 0.7) \times 10^{-5}$ pc$^{-3}$ in \citetalias{2011MNRAS.417...93R}.
However, the turnover of the halo luminosity function remains undetected similar to all previous works except \citetalias{2021MNRAS.502.1753T}, who claimed the discovery of the drop-off beyond $M_\mathrm{bol} \sim 15.5$, 
and will require deeper surveys to detect a turnover if exists.

\subsection{Faint white dwarfs}
\label{sec:faintWD}
For faint WDs with ${M}_{\mathrm{bol}}>16$, we observe a significantly larger number of objects (375 in total) compared to previous studies. In the following, we also discuss a subset of the WD candidates that have successfully passed a more stringent selection criterion based on the reduced chi-square ($\chi^2_\nu<1$); 
after applying this limit, we are left with 185 objects in our sample.
We will make an analysis of both samples in the following text.

\begin{figure}
\begin{center}
\includegraphics[width=\columnwidth]{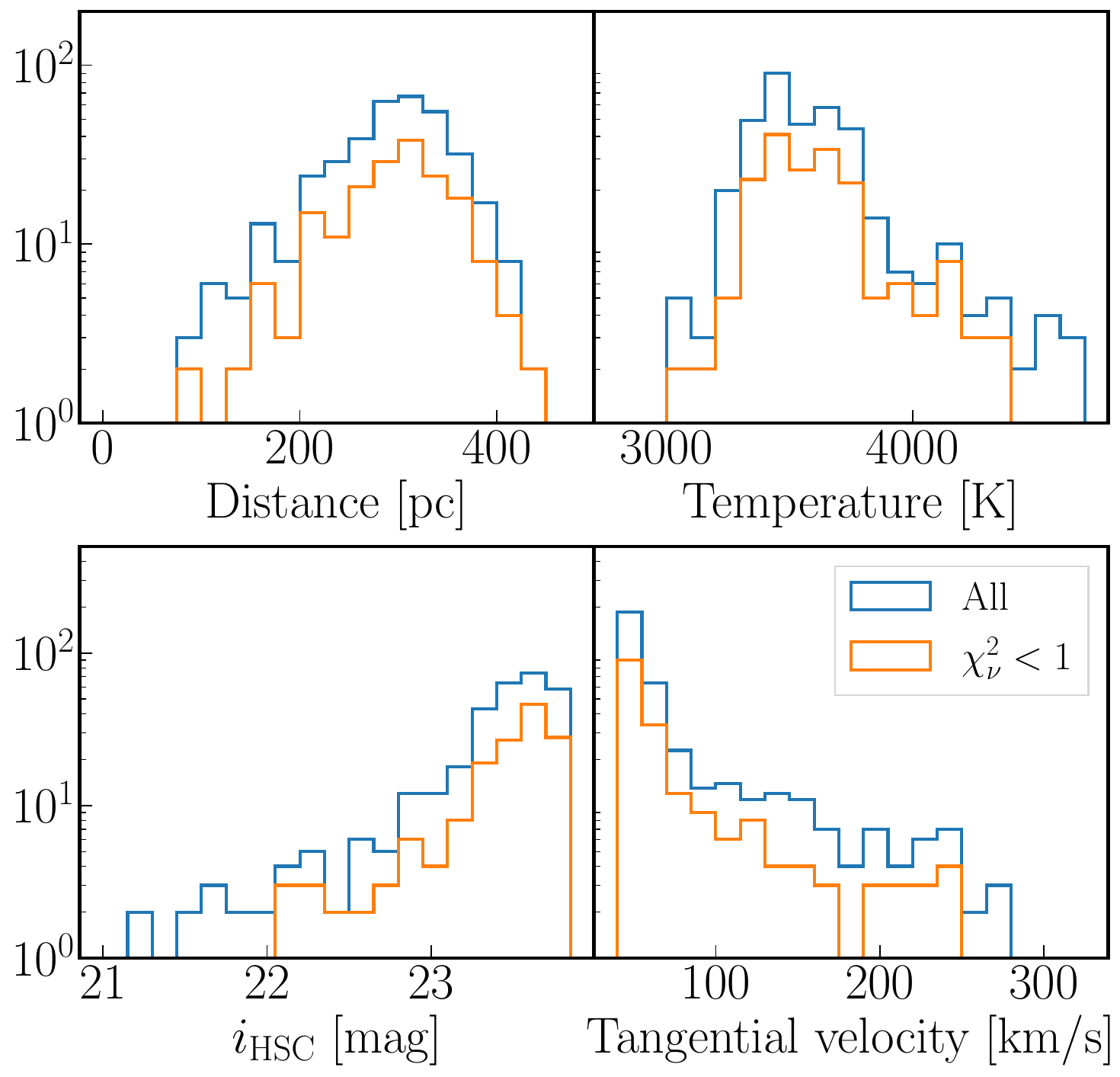}
 \caption{We display the distribution of the photometric distance, temperature and apparent magnitude in the $i$-band and tangential velocities, respectively, of the faint WD candidates.
 The blue histogram shows the results for the WD candidates with ${M}_{\mathrm{bol}}>16$, while the orange histogram shows the results for the subsample after a further cut $\chi^2_\nu<1$ for the blackbody fitting result. 
 }
 \label{fig:faintdist}
\end{center}
\end{figure}

In Figure~\ref{fig:faintdist}, the histograms display the distribution of these faint WDs across different distances, temperatures, apparent magnitudes and tangential velocity bins, respectively.
The abundance of WDs in the faint magnitude bins highlights the deep observational depth of our study, surpassing the capabilities of previous works.
Analysing the distribution of photometric distances for this faint WD sample, we find that the majority of these objects are concentrated at distances of approximately 300~pc. 
This is primarily attributed to the dominance of the thin disc population, which is characterised by its proximity to the Galactic plane. 
Consequently, most of the objects in this sample have tangential velocities that are just above the lower velocity limit of 40~km~s$^{-1}$. 
However, it is worth noting that a notable fraction of WDs with higher tangential velocities are also present in this sample. 
These objects likely originate from the stellar halo, which encompasses a more extended and dynamically hotter population of stars in the Galaxy. 
For the temperature distribution, the H and He atmosphere models are not applicable to temperatures below 3,000 K and are less reliable approaching the lower temperatures. 
This limitation arises from the lack of observations for WDs with such low temperatures. 
As a result, the blackbody model dominates the fitting for the WDs with lower temperatures,
though, the fitting for low-temperature WDs is still suffering from the goodness-of-fitting.

\begin{figure}
\begin{center}
 \includegraphics[width=\columnwidth]{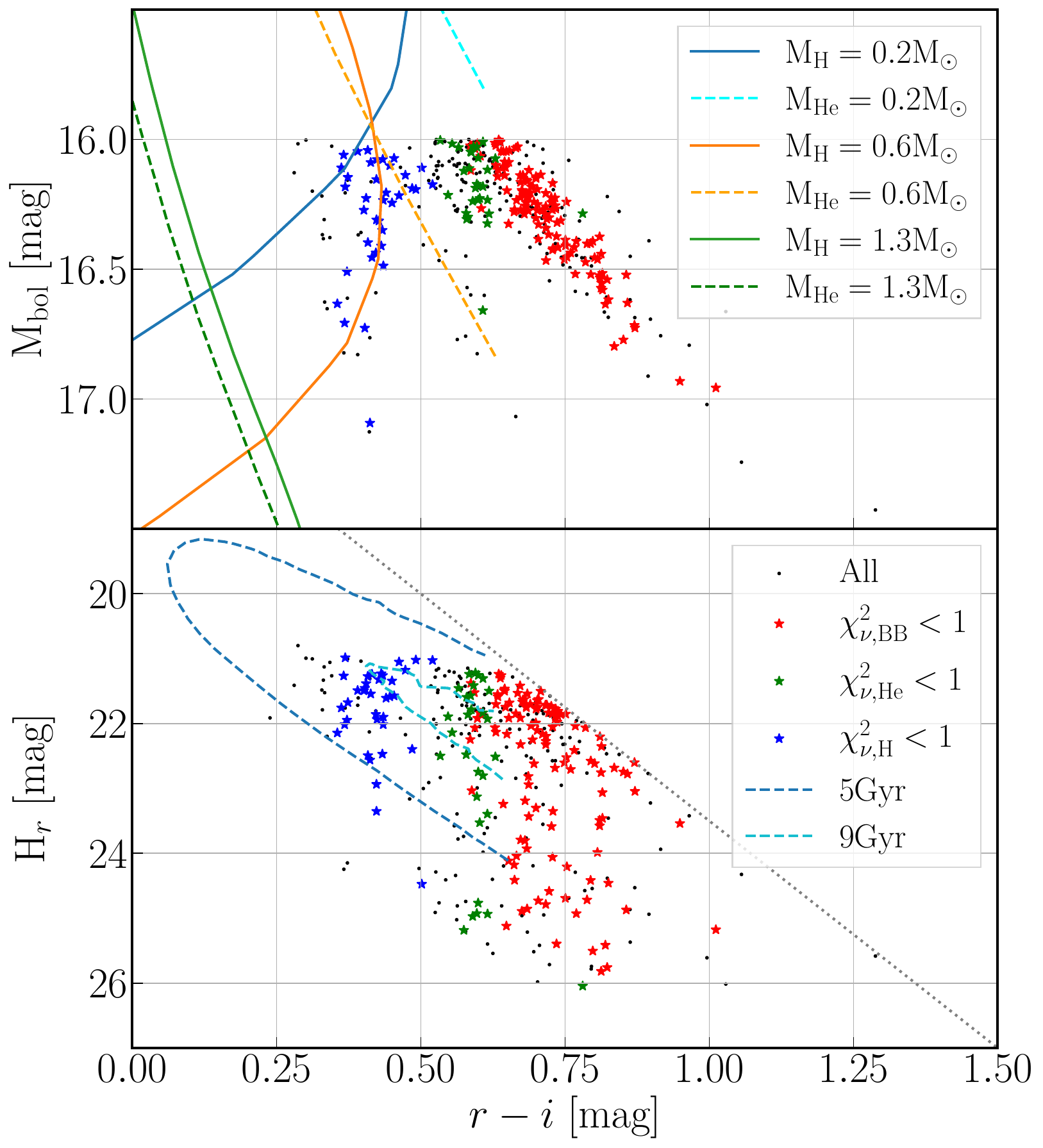}
 \caption{We show the bolometric magnitude and reduced proper motion versus colour $r-i$, respectively. 
 All the faint WD candidates are shown in black points. 
 The red stars represent the objects with a goodness-of-fit smaller than 1 while fitting to the blackbody model.
 The green and blue stars represent the He and H model as the legend shows.
 Both the H and He cooling sequences are depicted in the figure, with solid and dashed lines, respectively, representing different masses in different colours indicated in the legend. 
 The bottom panel illustrates the cooling-age contour, showing the cooling ages for a He model with a mass of 0.6 solar masses and a tangential velocity ranging from 40 km/s to 120 km/s.
 The contours shift downward as the tangential velocity increases, and these shifts are visually indicated by the shaded regions
 in corresponding colours. 
 The WDs fitted with a blackbody model are predominantly situated outside the shaded regions, 
 as these red and faint WDs lack sufficient observational data. 
 The current WD cooling sequences do not extend to such an old age range.
 }
 \label{fig:faintcdmrpm}
\end{center}
\end{figure}

In Figure~\ref{fig:faintcdmrpm}, the bolometric-colour diagram and the reduced proper motion diagram
showcase the faint WD sample.
We construct the theoretical lines for reference in the upper panel.
The solid line represents the models for H atmospheres, while the dashed line represents the models for He atmospheres. 
These models have been further differentiated by the inclusion of different masses as the legend shows.
The data points in the diagram correspond to the sample of faint WDs, with each point assigned a specific marker to indicate its best-fit model as the legend shows in the bottom panel.
The colour shows a distinguishable distribution for different models.
H-model WDs exhibit a bluer colour with higher temperatures, while He-model WDs tend to appear redder, aligning with observational data. 
However, older and fainter WDs, anticipated to be even redder, suffer from a dearth of observational coverage.
Our blackbody model is employed to characterise WDs with featureless spectra, indicative of objects that have undergone extensive burning out and consequently tend to exhibit a predominantly reddish colour.
In the bottom panel, the RPM diagram does not differ the cooling sequences from the variation from the mass or tangential velocity.
So we only show the cooling age contours derived from a specific He model assuming a 0.6$M_\odot$ and a tangential velocity ranging from 40~km~s$^{-1}$ to 120~km~s$^{-1}$.
By comparing the observed distribution of WDs in the RPM diagram to the age contours, we can briefly gain the age information of these faint WDs. 

The observed scatter in the properties of the faint WD candidates, particularly their cooling times 
exceeding 9~Gyr raises intriguing possibilities regarding their origin. 
As we mentioned in the previous section, these WDs are too old to have formed within our disc. 
Instead, they could have originated from other galaxies that merged with our Milky Way over billions of years.
The idea of extragalactic progenitors merging with our Galaxy has been explored in previous studies, such as the work by \cite{2018Natur.563...85H}. 
This study discussed the assembly history of the Milky Way through a merger with the Gaia-Enceladus (also known as Gaia sausage). 
The Gaia Mission found a large number of stars with elliptical orbits surrounding the central disc of the Milky Way Galaxy. 
These stars exhibited a different chemical composition compared to other stars in the halo, suggesting that they were not originally part of the Milky Way. 
Instead, they were believed to have originated from a dwarf galaxy, Gaia-Enceladus, which is thought to have collided with the Milky Way approximately 8 to 11 billion years ago. 
The stars from Gaia-Enceladus now cover nearly the entire sky, and their motions reveal the presence of streams and slightly retrograde and elongated trajectories. 
This collision and subsequent merger with Gaia-Enceladus had a significant impact on the Galactic evolution.
The merger of the Milky Way with Gaia-Enceladus resulted in the dynamical heating of the precursor of the Galactic thick disc. 
This event contributed to the formation of the thick disc component of the Milky Way approximately ten billion years ago \citep{2019NatAs...3..932G}. 
These findings are consistent with predictions from galaxy formation simulations, which suggest that the inner stellar halo of galaxies should be dominated by debris from only a few massive progenitors.
Previous work, as noted by \citetalias{2021MNRAS.502.1753T}, has also addressed the hypothesis regarding the burst of star formation attributed to the Gaia-Enceladus merger event.
This event is believed to have brought in ancient objects, including bright main-sequence stars and giants, with ages ranging from 10 to 13 Gyr.
Besides, the merger would also cause a period of intense star formation and contribute a large number of stars to the stellar halo and thick disc, which might also be the origin of those old WDs.
However, lack of additional information such as three-dimensional velocity and metallicity, we are still hard to say whether they are really from this structure or not.
If the presence of these old WD candidates in our sample is confirmed, it will add another piece to the puzzle of galactic evolution and merger events.

By relaxing the restriction to only $T < 4,000$~K instead of $M_{\mathrm{bol}} > 16$, we have identified a larger catalogue consisting of 699 objects. 
After applying a refinement based on the chi-square criterion, we are left with 398 objects. 
The determination of bolometric magnitudes relies on factors such as radius, which is obtained through techniques like MCMC or assumptions.
The reliability of the bolometric magnitudes is not as strong as the directly derived temperature values obtained from the spectra. 
Such a large sample of super cool WDs is unprecedented.
The work by \citet{2015MNRAS.449.3966G} listed 6 ultra-cool WDs with $T < 4,000$ K  and 54 cool WDs in total. 
They claimed a cooling age of 10~Gyr for these objects and provided a firm lower limit to the age of the thick disc population.

Overall, the potential existence of an ancient population of WDs, which may have formed during the early stages of our Galaxy, remains unconfirmed at this time. 
This extended catalogue of faint WD candidates serves as a valuable addition to the scarcity of observations for faint WDs with intriguing hints. 
Looking ahead, our next crucial step is to obtain spectroscopic observations for these candidates.
Confirmation of this unique population of WDs would not only contribute to our knowledge of the Galactic history but also shed light on the early stages of stellar evolution.

%%%%%%%%%%%%%%%%%%%%%%%%%%%%%%%%%%%%%%%%%%%%%%%%%%
\section{Conclusions}
\label{sec:conclusion}
In this paper, we have presented a comprehensive study of WDs in the Milky Way Galaxy, focusing on their selection, characterisation, and analysis. 
Using multi-band photometric data from the HSC survey and astrometric information from both the HSC and SDSS S82, we have discovered a large sample of WD candidates using the reduced proper motion diagram.
An important aspect of this work is the development and implementation of a robust selection algorithm 
to identify WD candidates for the construction of the luminosity function.
Through careful calibration and validation, we have established a selection method that effectively separates WDs from other stellar populations, minimising contamination and maximising the purity of our samples. 
On the other hand, the algorithm also incorporates proper motion limits, photometric criteria, and goodness-of-fit thresholds to ensure the reliability of the photometric estimate of the distance and bolometric magnitude by fitting to the blackbody spectrum and WD atmosphere models.
We found 5,080 WD candidates from our selection criteria in the HSC $i$-band magnitude range from 19 to 24, covering 165~deg$^2$ of the sky.
 
We successfully constructed the luminosity functions for the Galactic disc and halo WD candidates. 
By considering different tangential velocity ranges, we were able to investigate the contributions of different stellar populations to the overall LF. 
The number densities obtained were found to be $(0.945 \pm 0.094) \times 10^{-2}$~pc$^{-3}$ and $(4.20 \pm 1.74) \times 10^{-4}$~pc$^{-3}$, respectively. 
Furthermore, we still compared our results with those of previous studies.
Our findings revealed significantly higher number densities in both the disc and halo components compared to previous work, which can be attributed to the unprecedented depth of our observations. 
These results demonstrate the importance of deep observations and the application of advanced selection algorithms in unveiling the properties and characteristics of faint WDs in the Milky Way. 

We have discovered a sample of 677 WD candidates with temperatures lower than 4,000 K, which is an unprecedentedly large number of faint WDs. 
Using a rough estimation of the photometric distance to calculate the absolute magnitude, a total of 375 WD candidates have been identified with bolometric magnitudes fainter than 16. 
This result is particularly intriguing, as it implies the presence of a substantial number of faint WDs in the population.
The exploration of the intrinsically faint sample of WDs in this study suggests the possibility of uncovering a remarkably enigmatic population: the first generation of WDs.
Our findings of the potentially oldest WDs could provide clues to unravel the fossils of the merging events that triggered the active assembly history of the Milky Way. 
If we can spectroscopically confirm the existence of these WDs, it would give us a glimpse into the distant past of our Galaxy. 
We plan to explore this exciting direction for our future work. 

%%%%%%%%%%%%%%%%%%%%%%%%%%%%%%%%%%%%%%%%%%%%%%%%%%
\section*{Acknowledgements}
The HSC collaboration includes the astronomical communities of Japan and Taiwan, as well as Princeton University. 
The HSC instrumentation and software were developed by the National Astronomical Observatory of Japan (NAOJ), 
the Kavli Institute for the Physics and Mathematics of the Universe (Kavli IPMU), 
the University of Tokyo, 
the High Energy Accelerator Research Organization (KEK), 
the Academia Sinica Institute for Astronomy and Astrophysics in Taiwan (ASIAA), 
and Princeton University. 
Funding was contributed by the FIRST program from the Japanese Cabinet Office, 
the Ministry of Education, Culture, Sports, Science and Technology (MEXT), 
the Japan Society for the Promotion of Science (JSPS), 
Japan Science and Technology Agency (JST), 
the Toray Science Foundation, NAOJ, Kavli IPMU, KEK, ASIAA, and Princeton University.
This work is supported in part by the World Premier International
Research Center Initiative (WPI Initiative), MEXT, JSPS
KAKENHI Grant Numbers JP19H00677, JP20H05850, JP20H05855, JP22J11959, JP20K04010, JP20H01904, JP22H0013, JP24H00215, and
Basic Research Grant (Super AI) of Institute for AI and 
Beyond of the University of Tokyo. 

This work is based on data collected at the Subaru Telescope and retrieved from the HSC data archive system, which is operated by the Subaru Telescope and Astronomy Data Center (ADC) at NAOJ. 
Data analysis was in part carried out with the cooperation of the Center for Computational Astrophysics, NAOJ.
We are honoured and grateful for the opportunity to observe the Universe from Maunakea, which has cultural, historical and natural significance in Hawaii.

This work makes use of software developed for the Vera C. Rubin Observatory. 
We thank the Rubin Observatory for making their code available as free software at \url{http://pipelines.lsst.io/}.

The Pan-STARRS1 Surveys (PS1) and the PS1 public science archive have been made possible through contributions by the Institute for Astronomy, the University of Hawaii, the Pan-STARRS Project Office, the Max Planck Society and its participating institutes, 
the Max Planck Institute for Astronomy, Heidelberg, and the Max Planck Institute for Extraterrestrial Physics, Garching, 
the Johns Hopkins University, 
Durham University, 
the University of Edinburgh, the Queen’s University Belfast, 
the Harvard-Smithsonian Center for Astrophysics, 
the Las Cumbres Observatory Global Telescope Network Incorporated, the National Central University of Taiwan, the Space Telescope Science Institute, the National Aeronautics and Space Administration under grant No. NNX08AR22G issued through the Planetary Science Division of the NASA Science Mission Directorate, the National Science Foundation grant No. AST-1238877, the University of Maryland, Eotvos Lorand University (ELTE), the Los Alamos National Laboratory, and the Gordon and Betty Moore Foundation.

We also make use of \texttt{astropy}, a community-developed \texttt{PYTHON} package for Astronomy \citep{2013A&A...558A..33A} and \texttt{WDPhotTools}, a WD toolkit in \texttt{PYTHON} \citep{2022RASTI...1...81L}.

%%%%%%%%%%%%%%%%%%%%%%%%%%%%%%%%%%%%%%%%%%%%%%%%%%

%%%%%%%%%%%%%%%%%%%%%%%%%%%%%%%%%%%%%%%%%%%%%%%%%%
\section*{Data Availability}
The data underlying this article are available in HSC-SSP, at \url{https://doi.org/10.1093/pasj/psab122}, SDSS DR7 Stripe82 at \url{https://doi.org/10.1088/0004-637X/794/2/120}, Gaia DR3 at \url{https://doi.org/10.1051/0004-6361/202243940}, and SDSS DR14 quasar at \url{https://doi.org/10.1051/0004-6361/201732445}. 
The datasets were derived from sources in the public domain: HSC-SSP (\url{https://hsc-release.mtk.nao.ac.jp/doc/}), SDSS Catalogue Archive Server (\url{http://skyserver.sdss.org/CasJobs/}), Gaia Archive (\url{https://gea.esac.esa.int/archive/}), and SDSS Quasar Catalogue (\url{https://www.sdss4.org/dr17/algorithms/qso_catalog/}).


\bibliographystyle{mnras}
\bibliography{ref} % if your bibtex file is called example.bib

%\begin{thebibliography}{99}
%\end{thebibliography}

%%%%%%%%%%%%%%%%%%%%%%%%%%%%%%%%%%%%%%%%%%%%%%%%%%

%%%%%%%%%%%%%%%%% APPENDICES %%%%%%%%%%%%%%%%%%%%%

%%%%%%%%%%%%%%%%%%%%%%%%%%%%%%%%%%%%%%%%%%%%%%%%%%

% Don't change these lines
\bsp	% typesetting comment
\label{lastpage}
\end{document}